\documentclass[sigconf,screen]{acmart}
\renewcommand\footnotetextcopyrightpermission[1]{} 

\usepackage{cleveref} 
\usepackage{xspace} 
\usepackage{multirow}
\usepackage{makecell}
\usepackage{array} 
\usepackage{newtxtext,newtxmath}
\usepackage{booktabs}
\usepackage{float}

\usepackage{listings}
\usepackage{xcolor} 
\crefformat{figure}{Figure~#2#1#3}
\crefformat{table}{Table~#2#1#3}
\crefformat{equation}{Equation~#2#1#3}
\crefformat{section}{Section~#2#1#3}

\begin{document}

\title{ControlFoley: Unified and Controllable Video-to-Audio Generation with Cross-Modal Conflict Handling}

\author{Jianxuan Yang}
\authornote{Equal contribution.}
\authornote{Corresponding Author. Email: yangjianxuan@xiaomi.com}
\affiliation{
  \institution{MiLM Plus, Xiaomi Inc.}
  \country{}
}

\author{Xinyue Guo}
\authornotemark[1]
\affiliation{
  \institution{MiLM Plus, Xiaomi Inc.}
  \country{}
}

\author{Zhi Cheng}
\affiliation{
  \institution{MiLM Plus, Xiaomi Inc.}
  \institution{Wuhan University}
  \country{}
}

\author{Kai Wang}
\affiliation{
  \institution{MiLM Plus, Xiaomi Inc.}
  \institution{Wuhan University}
  \country{}
}

\author{Lipan Zhang}
\affiliation{
  \institution{MiLM Plus, Xiaomi Inc.}
  \country{}
}

\author{Jinjie Hu}
\affiliation{
  \institution{MiLM Plus, Xiaomi Inc.}
  \country{}
}

\author{Qiang Ji}
\affiliation{
  \institution{MiLM Plus, Xiaomi Inc.}
  \country{}
}

\author{Yihua Cao}
\affiliation{
  \institution{MiLM Plus, Xiaomi Inc.}
  \country{}
}

\author{Yihao Meng}
\affiliation{
  \institution{MiLM Plus, Xiaomi Inc.}
  \institution{Wuhan University}
  \country{}
}

\author{Zhaoyue Cui}
\affiliation{
  \institution{MiLM Plus, Xiaomi Inc.}
  \institution{Wuhan University}
  \country{}
}

\author{Mengmei Liu}
\affiliation{
  \institution{MiLM Plus, Xiaomi Inc.}
  \country{}
}

\author{Meng Meng}
\affiliation{
  \institution{MiLM Plus, Xiaomi Inc.}
  \country{}
}

\author{Jian Luan}
\affiliation{
  \institution{MiLM Plus, Xiaomi Inc.}
  \country{}
}


\renewcommand{\shortauthors}{Yang et al.}

\begin{abstract}

Recent advances in video-to-audio (V2A) generation have enabled high-quality audio synthesis from visual content, yet achieving robust and fine-grained controllability remains a fundamental challenge. In particular, existing methods suffer from two key limitations: weak textual controllability under visual-text semantic conflict, and imprecise stylistic control due to entangled temporal and timbre information in reference audio. Moreover, the lack of standardized benchmarks further hinders systematic evaluation of controllability.

In this paper, we propose ControlFoley, a unified and controllable multimodal V2A framework that enables precise control across video, text and reference audio. First, we introduce a joint visual encoding paradigm that integrates CLIP with a spatio-temporal audio-visual encoder, enhancing both audio-visual alignment and textual controllability under cross-modal conflict. Second, we propose a temporal-timbre decoupling strategy that suppresses redundant temporal information in reference audio while preserving discriminative timbre features, enabling accurate and interference-free stylistic control. Third, we design a modality-robust training scheme with unified multimodal representation alignment (REPA) and random modality dropout, aligning generated audio representations with aggregated multimodal conditions. Finally, we present VGGSound-TVC, the first benchmark specifically designed to quantify textual controllability under varying degrees of visual-text semantic conflict.

Extensive experiments demonstrate that ControlFoley achieves state-of-the-art performance across multiple V2A tasks, including text-guided, text-controlled, and audio-controlled generation, outperforming both task-specific and unified baselines. 
Furthermore, ControlFoley exhibits superior controllability under cross-modal conflict while maintaining strong temporal synchronization and audio quality. 
It also demonstrates competitive or superior performance compared to an industrial V2A system, establishing a unified framework for controllable V2A generation.

Codes, pretrained models, datasets and demos are available at: \url{https://github.com/xiaomi-research/controlfoley}.

\end{abstract}



\keywords{ Video-to-Audio Generation, Multimodal Generation, Controllable Generation, Multimodal Learning, Cross-Modal Conflict, Audio Generation}
\begin{teaserfigure}
  \includegraphics[width=\textwidth]{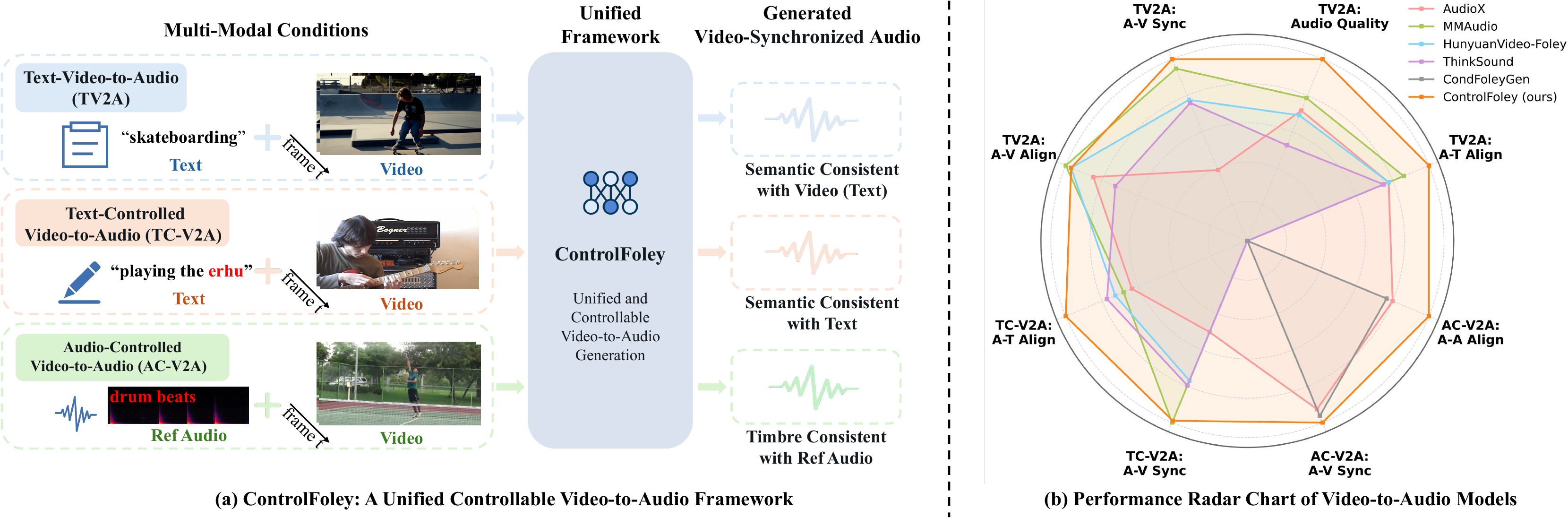}
  \caption{Left: Overview of the ControlFoley framework with three multimodal conditioning modes for controllable video-synchronized audio generation. Right: Performance radar chart of Video-to-Audio models.}
  \Description{teaser figure.}
  \label{fig:teaser}
\end{teaserfigure}


\maketitle

\section{Introduction}
In recent years, the rapid advancement of multimodal generative models has revolutionized content creation workflows, enabling efficient, cost-effective audio dubbing for film, games, and advertising. Representative Video-to-Audio (V2A) works such as MMAudio, HunyuanVideo-Foley, Kling-Foley and ThinkSound have demonstrated strong foundational capabilities, effectively bridging silent visual content with synchronized audio ~\cite{zhang2024foleycrafter, Cheng2025MMAudio, shan2025hunyuanvideofoley, liu2025thinksound, wang2025klingfoley,wang2025audiogenomni}. However, controllable and interactive audio generation remains a largely unresolved challenge. This is especially true in scenarios requiring high customization and creative expression, such as crafting a unique stylized soundscape for a film or generating immersive environmental audio for a game. To unlock this creative potential, V2A systems must evolve beyond passive audio-visual mapping to incorporate robust multimodal guidance, empowering creators to precisely shape outputs according to their artistic intent.

Existing multimodal controlled V2A methods have made initial strides toward this objective, yet their architectural designs exhibit inherent limitations that hinder robust, fine-grained control. As an early pioneer in audio-conditioned V2A, CondFoleyGen ~\cite{du2023condfoleygen} introduces analogy-based timbre transfer via a Transformer-VQGAN pipeline, specializing in audio-visual paired conditioning. This task-focused design limits flexibility in diverse multimodal scenarios, while its onset-aligned transfer logic struggles with complex actions or ambiguous temporal cues. MultiFoley ~\cite{Chen2025Multifoley} extends to simultaneous text and reference audio control via a diffusion transformer, aiming for audio attribute disentanglement. Yet its low frame rate visual features for temporal alignment degrade synchronization and destabilize cross-modal fusion. As a representative unified multimodal "anything-to-audio" framework, AudioX ~\cite{tian2026audiox} adopts a generic Multimodal Adaptive Fusion (MAF) module to integrate text, video, and reference audio inputs. Lacking fine-grained regulatory mechanisms, this design leads to limited control robustness and suboptimal temporal synchronization performance. Despite these advancements, fundamental challenges in the design of effective control mechanisms for V2A generation remain largely unaddressed.

Such control deficiencies are particularly pronounced in the two dominant V2A control paradigms. For Text-Controlled V2A (TC-V2A), the textual modality often plays a marginal role in generation control: when textual descriptions are semantically consistent with visual content, text merely acts as supplementary detail to refine minor audio characteristics. In contrast, in creative stylization scenarios requiring semantic divergence between text and vision, a prevalent "visual dominance" phenomenon emerges. Under such semantic conflict conditions, textual guidance is consistently overridden by salient visual cues in cross-modal fusion, resulting in control failure where the generated audio still predominantly aligns with visual content rather than textual instructions. A further compounding issue is the lack of a dedicated benchmark for rigorously evaluating generation performance under varying degrees of text-visual semantic conflict. Consequently, despite the rapid proliferation of TC-V2A methods, their actual controllability remains difficult to objectively quantify and compare.

While text-based control offers inherent semantic flexibility, it is inherently limited in specifying highly personalized or unique audio characteristics---e.g., the bespoke vocalization of a fictional creature or the distinctive sound of a magical effect---that are nearly impossible to describe with precise linguistic expressions. To address this limitation, Audio-Controlled V2A (AC-V2A) has emerged as a powerful complementary paradigm. By leveraging reference audio clips as conditional guidance, AC-V2A enables fine-grained control over audio timbre and style, facilitating the generation of highly personalized audio that transcends the expressiveness of textual descriptions. The core objective of AC-V2A is to generate audio that maintains strict temporal synchronization with visual content while faithfully inheriting the acoustic properties of the reference audio. This paradigm, however, introduces a core technical challenge: reference audio inherently contains both the desired timbre features and redundant native temporal information. As evidenced by the shortcomings of prior work---such as CondFoleyGen’s onset-aligned transfer, MultiFoley’s lack of explicit timbre-temporal disentanglement---existing methods fail to effectively decouple timbre and temporal information in reference audio, frequently leading to temporal interference and imprecise stylistic control.

To address these multifaceted challenges, we propose ControlFoley, a unified multimodal V2A model with four key innovations:
\begin{itemize}
\item \textbf{Joint Visual Encoding with CAV-MAE-ST for Enhanced Textual Controllability}: We propose a joint visual encoding paradigm that combines a spatio-temporally optimized CAV-MAE-ST encoder with CLIP visual features. CAV-MAE-ST strengthens fine-grained audio-visual alignment and cross-modal semantics, while the dual-branch paradigm consisting of CAV-MAE-ST and CLIP mitigates text-visual conflict and enhances textual control authority in TC-V2A.
\item \textbf{Temporal-Timbre Decoupling for Precise Timbre Control}: Built upon the MMDiT-based V2A framework, we design a timbre-focused control strategy for reference audio---retaining only timbre information and suppressing redundant temporal information to avoid temporal interference. This strategy enables precise timbral guidance, facilitating accurate timbre control for AC-V2A tasks.
\item \textbf{Modality-Robust Training with Unified REPA}: We ensure robust performance under arbitrary modality absence, allowing ControlFoley to excel at all multimodal controlled generation tasks. Specifically, we apply random dropout to all modalities during training; additionally, we aggregate the available visual, textual, and reference audio representations as the target representation to align with the generated audio features. This design enhances DiT’s alignment with high-level features across all modalities, thereby boosting overall generation performance.
\item \textbf{VGGSound-TVC: A Benchmark to Fill the Evaluation Gap in Text-Controlled V2A}: We construct the VGGSound-TVC dataset, the first dedicated benchmark for evaluating textual controllability and generation performance under varying degrees of visual-text semantic conflict. This dataset fills the gap in standardized evaluation for TC-V2A tasks.
\end{itemize}

Extensive experiments demonstrate that ControlFoley achieves state-of-the-art performance across three core V2A tasks (TV2A, TC-V2A, AC-V2A), while significantly improving controllability and robustness under challenging multimodal conditions.

\begin{figure*}[t]
  \centering
  \includegraphics[width=0.9\linewidth]{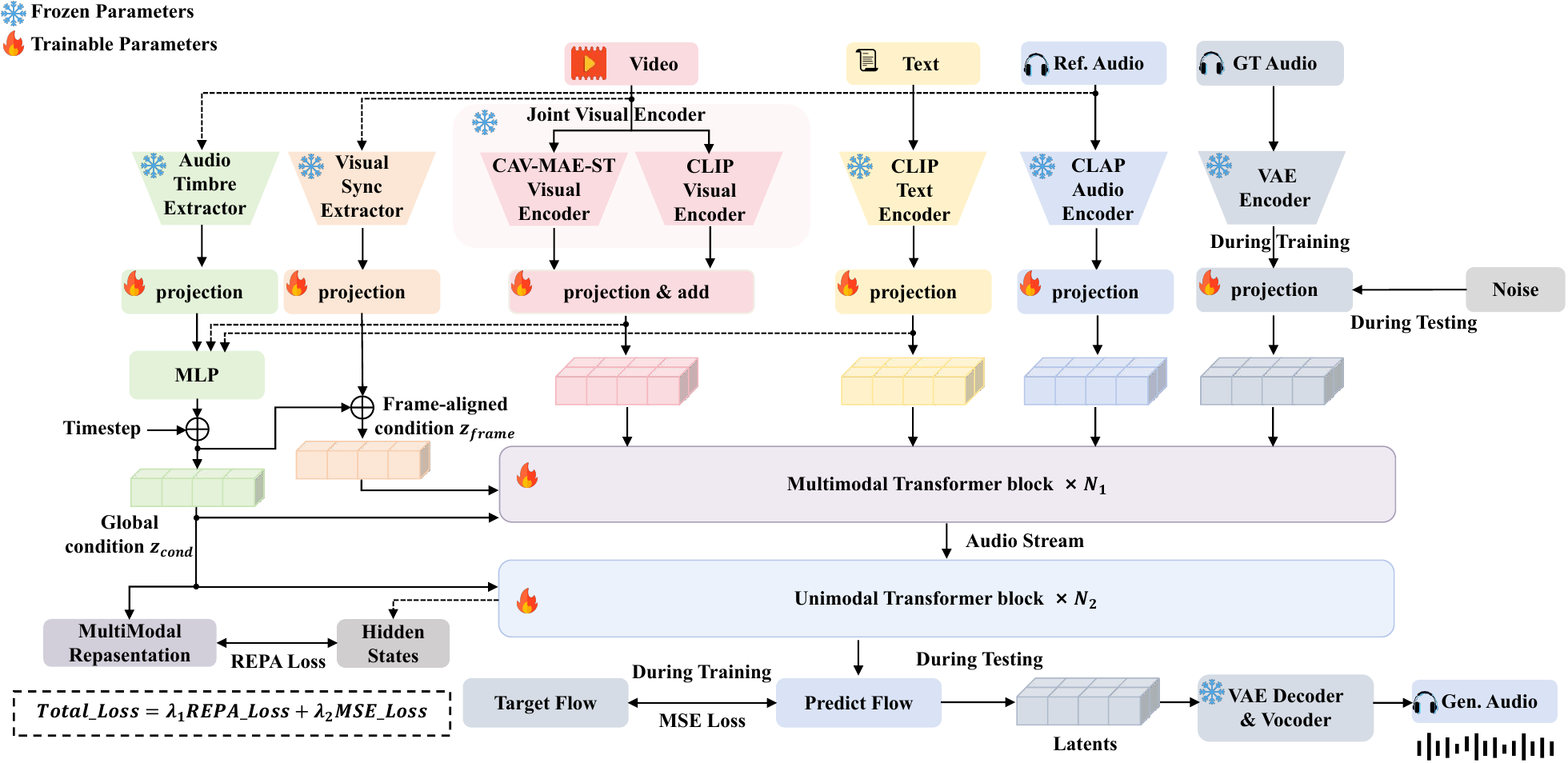}
  \caption{Overview of the ControlFoley model architecture. The proposed model integrates visual, text, and audio features into a multimodal transformer network with $N_1$ multimodal DiT blocks and $N_2$ unimodel DiT blocks. Global conditions incorporating visual semantics, textual semantics, and timbre features, together with frame-aligned synchronization conditions, are integrated into the network for precise semantics and synchronization control.}
  \label{fig:controlfoley}
\end{figure*}

\section{Related Works}

\subsection{Audio-Visual Representation Learning}
Robust audio-visual representation learning is pivotal for high-quality video-to-audio (V2A) generation, as it underpins precise temporal synchronization and semantic alignment between visual content and synthesized audio ~\cite{iashin2024synchformer, liu2025valor, luo2023difffoley, Li2025FCConDubber, tsiamas2025sequential, viertola2025vaura}. Masked signal modeling has emerged as a potent approach for capturing fine-grained cross-modal features: it compels encoders to infer intrinsic semantic and structural information by reconstructing masked segments of audio or visual inputs. State-of-the-art approaches ~\cite{gong2023cavmae, Georgescu2023avmae, Araujo2025cavmaesync} have advanced this paradigm by integrating masked autoencoding with contrastive learning, enabling the joint capture of modality-specific details and cross-modal correlations to yield robust joint representations for generative tasks.

Building on this hybrid framework, CAV-MAE-Edit ~\cite{guo2026avedit} tailors audio-visual learning for sound effect editing via frame segmentation and global token aggregation, refining temporal matching and global semantic capture. It further introduces audio mixing as a training strategy---mixing target and irrelevant audio to enhance the model’s audio disentanglement capability for editing-centric tasks. While effective for sound editing, this design is ill-suited for pure V2A generation: it introduces non-essential computational overhead and diverts the model’s focus from the core demand of precise audio-visual alignment for novel audio synthesis.

\subsection{Multimodal Controlled Video-to-Audio Generation}
The pursuit of controllable audio synthesis has driven the evolution of V2A models from single-condition generators to sophisticated multimodal frameworks. These advanced systems aim to provide creators with fine-grained control over the generated audio by incorporating guidance from various modalities, including text and reference audio. We categorize and review them based on their primary control modality.

\noindent\textbf{Text-Guided Generation.} Text serves as a flexible control signal for audio semantic customization ~\cite{hung2025tangoflux, liu2024audioldm2,haji2026genauL}, leading to two overlapping yet distinct research directions based on textual control authority. Most mainstream works fall under Text-Video-to-Audio (TV2A), which prioritizes audio-video alignment while integrating text as a supplementary constraint to refine audio attributes. Representative models like MMAudio ~\cite{Cheng2025MMAudio} and HunyuanVideo-Foley ~\cite{shan2025hunyuanvideofoley} incorporate text into multimodal fusion but lack dedicated mechanisms to prioritize textual guidance---their textual guidance primarily functions as a semantic filter, refining video-aligned audio rather than dictating its core content. Foley Control ~\cite{rowles2025foleycontrol} offers a parameter-efficient alternative by connecting frozen V-JEPA2 and text-to-audio backbones through a lightweight cross-attention bridge, allowing text to specify global semantics while video refines local timing.

A smaller line of research targets Text-Controlled V2A (TC-V2A), where text is intended as the dominant control signal. Early explorations like FoleyCrafter ~\cite{zhang2024foleycrafter} attempted to steer audio semantics via text, but its lightweight design struggles to override visual dominance. More recently, unified frameworks like AudioX ~\cite{tian2026audiox} extended text control capabilities across modalities, yet its generic fusion module limits robust textual authority under conflict.

A critical gap persists: Existing V2A models have weak textual control, failing to deliver robust semantic customization under text-visual conflict. Compounded by the absence of standardized benchmarks for textual controllability, the core challenge of "reliable textual control while maintaining audio-video synchronization" remains unaddressed.

\noindent\textbf{Reference Audio-Guided Generation.} To enable personalized stylistic control (e.g., timbre transfer) beyond the expressiveness of text ~\cite{garcia2025sketch2sound}, Audio-Controlled V2A (AC-V2A) incorporates reference audio as a core conditional input. Early specialized models, such as CondFoleyGen ~\cite{du2023condfoleygen}, demonstrated the feasibility of this paradigm. It adopts a Transformer-VQGAN architecture, leveraging self-supervised intra-video training to transfer timbre while aligning with video onsets. However, its task-specialized design limits flexibility in multimodal scenarios, and its onset-aligned transfer logic struggles with complex actions or ambiguous temporal cues.

A key trend thereafter shifted toward unified frameworks supporting multiple control modalities. AudioX ~\cite{tian2026audiox} integrates reference audio into its "anything-to-audio" framework via a generic Multimodal Adaptive Fusion (MAF) module, achieving strong instruction-following capabilities but lacking robust control over stylistic attributes due to its one-size-fits-all fusion design. MultiFoley ~\cite{Chen2025Multifoley} represents a notable advancement, supporting reference audio alongside text and audio extension via a diffusion transformer framework. It achieves attribute separation via multiconditional training on diverse datasets. However, it fails to implement explicit timbre-temporal disentanglement---its unified audio latent representation retains redundant temporal dynamics from reference audio, leading to temporal interference and imprecise stylistic control in multi-event scenarios.

A critical gap persists in AC-V2A: existing methods either lack control flexibility or fail to explicitly isolate desired stylistic features. The core challenge of "cleanly decoupling timbre from redundant temporal information in reference audio" remains unaddressed, hindering precise and robust stylistic control.

\noindent\textbf{Our work.} Building on the aforementioned advancements, we propose ControlFoley, a unified multimodal V2A framework that directly addresses the core limitations of existing methods in textual control, stylistic generation and multimodal robustness, while filling the evaluation gap in the field.

For textual control, unlike the former that lack textual guidance prioritization mechanisms and fail to resolve text-visual conflicts, we design a joint visual encoding strategy to enhance textual control dominance while preserving precise audio-video alignment, bridging the gap between weak text supplementation and desired strong TC-V2A capability.

For AC-V2A stylistic generation, we advance beyond prior methods: we propose an explicit Temporal-Timbre Decoupling strategy that discards redundant temporal dynamics from reference audio while retaining discriminative timbre features. This enables interference-free precise stylistic control in our unified V2A framework.

For multimodal robustness, we devise a unified supervision mechanism combining random modality dropout and REPA loss with integrated visual-text-timbre feature alignment to ensure stable high-quality generation under arbitrary modality absence.

Finally, to address the lack of standardized textual controllability evaluation, we construct VGGSound-TVC---the first dedicated benchmark for quantifying textual control performance under varying text-visual semantic conflict, filling a critical evaluation gap and providing a rigorous unified protocol for the community.

\section{Method}

\subsection{Overview}

Our framework builds upon a multimodal diffusion transformer (MMDiT) ~\cite{Esser2024ScalingRF,Hoogeboom2023simpdiff, Cheng2025MMAudio} backbone for controllable video-to-audio generation. Based on this architecture, we introduce several key modifications to improve multimodal alignment and controllability in V2A generation. First, we propose a \textbf{Joint Visual Encoding Paradigm} that combines a CLIP visual encoder ~\cite{radford2021clip} with the proposed CAV-MAE-ST encoder to jointly capture vision-language and audio-visual correlations. This design provides complementary representations that enhance cross-modal synergy while mitigating modality competition between visual and textual inputs, thereby improving the controllability of generated audio with respect to textual instructions. Second, we extend the backbone with a \textbf{Timbre-Focused Reference Audio Control} mechanism, enabling the model to incorporate reference audio for acoustic style control. Specifically, reference audio is introduced through a dedicated conditioning branch in the multimodal transformer. To ensure that the reference signal conveys acoustic style rather than temporal dynamics, the branch is designed to suppress temporal modeling. In addition, a pretrained timbre encoder extracts timbre representations that are injected as global conditioning signals. Third, we employ a \textbf{Multimodal REPA Loss} to further improve multimodal alignment during training. Finally, to systematically evaluate textual controllability under visual-text semantic conflict, we introduce a new benchmark named VGGSound-TVC, which enables quantitative analysis of modality dominance under different levels of visual-text inconsistency.

An overview of the proposed framework is illustrated in \cref{fig:controlfoley}.

\subsection{Joint Visual Encoding with CAV-MAE-ST}

\begin{figure}[t]
  \centering
  \includegraphics[width=\linewidth]{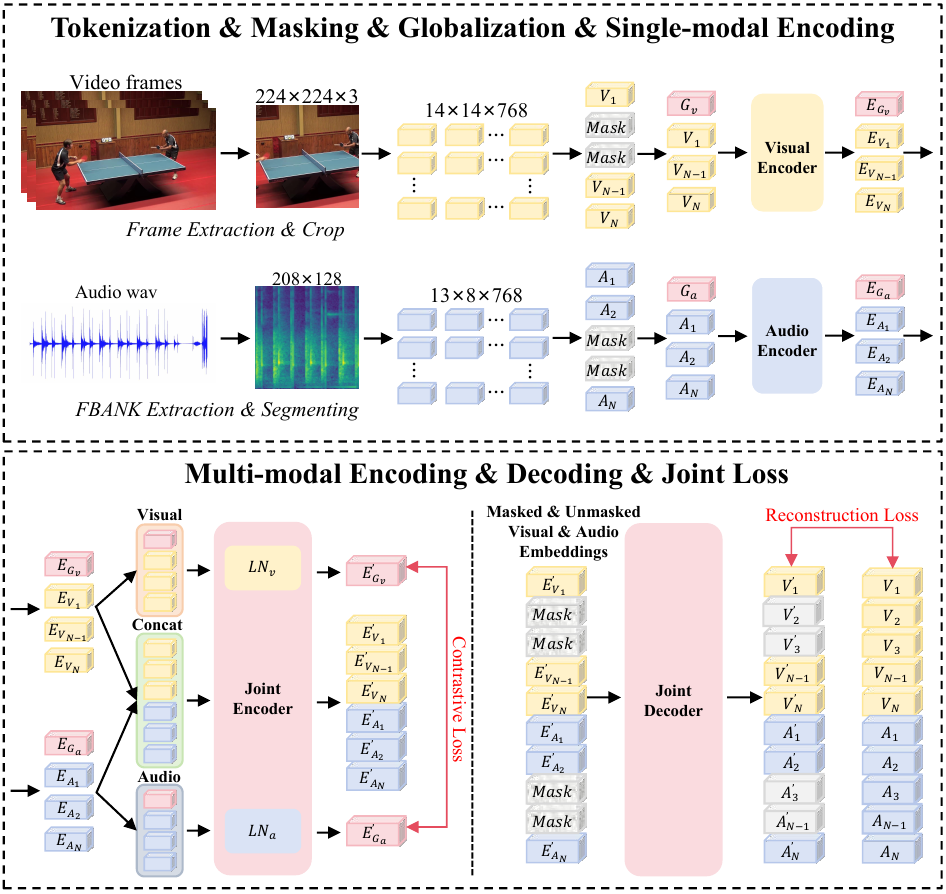}
  \caption{Architecture of the proposed CAV-MAE-ST. The model adopts masked audio-visual learning with spatio-temporal tokenization. Compared to CAV-MAE-Edit, editing-specific components such as audio mixing are removed.}
  \label{fig:cav-mae-st}
\end{figure}

\textbf{CAV-MAE-ST Visual Encoder.} Robust audio-visual representation learning is essential for V2A generation, as it underpins accurate temporal synchronization and semantic alignment between visual events and synthesized audio. To this end, we build upon the masked audio-visual modeling paradigm of CAV-MAE~~\cite{gong2023cavmae}, and propose a streamlined spatio-temporal encoder, termed CAV-MAE-ST, tailored for V2A generation.

As illustrated in \cref{fig:cav-mae-st}, given an input video $V$ and its corresponding audio $A$, we first perform modality-specific tokenization:
\begin{equation}
\tilde{z}_v = \{v_i\}_{i=1}^N = T_v(V), \quad
\tilde{z}_a = \{a_i\}_{i=1}^N = T_a(A),
\end{equation}
where $T_v(\cdot)$ and $T_a(\cdot)$ denote the visual and audio tokenization processes. Specifically, $T_v$ performs frame extraction and spatial patchification to produce spatio-temporal visual tokens, while $T_a$ converts audio waveforms into log-Mel spectrograms and segments them into temporal audio tokens, as shown in \cref{fig:cav-mae-st}. Additionally, a learnable global token (denoted as $g_v$ for visual and $g_a$ for audio) is appended to capture global context.

The tokenized features are then encoded by modality-specific encoders:
\begin{equation}
z_v = \{e_{v_i}\}_{i=1}^N = E_v(\tilde{z}_v), \quad
z_a = \{e_{a_i}\}_{i=1}^N = E_a(\tilde{z}_a),
\end{equation}
where $E_v(\cdot)$ and $E_a(\cdot)$ denote the visual and audio encoders, respectively. The global tokens are encoded in the same manner to produce global representations.

Following the masked audio-visual modeling paradigm, a subset of tokens from both modalities is randomly masked. A shared multimodal encoder is used to reconstruct the masked tokens from cross-modal context, with the reconstruction objective defined as:
\begin{equation}
\mathcal{L}_{\mathrm{rec}} =
\|z_v - \hat{z}_v\|_2^2 +
\|z_a - \hat{z}_a\|_2^2 .
\end{equation}

Compared to the original CAV-MAE, we introduce a spatio-temporal (ST) modeling strategy to better capture temporal dynamics critical for V2A generation. In particular, we adopt a frame-level segmentation scheme that preserves temporal ordering across video frames, enabling the model to learn fine-grained motion patterns and their corresponding acoustic events, as reflected by the temporal tokenization in \cref{fig:cav-mae-st}.

We further draw inspiration from CAV-MAE-Edit, which introduces editing-oriented components such as audio mixing to facilitate source disentanglement. However, such designs are not aligned with the objective of V2A generation, where the goal is to synthesize temporally synchronized audio rather than separate mixed sources. Therefore, as shown in \cref{fig:cav-mae-st}, we remove editing-specific modules and retain a simplified masked reconstruction pipeline, allowing the model to focus on learning precise audio-visual correspondence.

Overall, CAV-MAE-ST preserves the strengths of masked audio-visual learning while incorporating temporal modeling and eliminating task-misaligned components. The resulting representations exhibit improved temporal sensitivity and stronger audio-visual alignment, providing a solid visual foundation for the subsequent multimodal generation framework.

\noindent\textbf{Joint Visual Encoding Paradigm.} Most existing V2A generation models employ visual encoders that are aligned with textual representations through large-scale vision-language pretraining. Depending on whether visual and textual representations share the same backbone, current approaches typically adopt either a unified encoder (e.g., CLIP) ~\cite{Cheng2025MMAudio} or separate encoders that still rely on vision-language aligned visual features ~\cite{tian2026audiox, shan2025hunyuanvideofoley, liu2025thinksound}.

However, such designs introduce two limitations for controllable V2A generation. First, visual representations aligned solely with text may overlook intrinsic audio-visual correlations that are essential for realistic sound generation, such as motion-induced acoustic events and temporal synchronization between visual actions and sounds. Second, the interaction between visual and textual features becomes unstable when their semantics differ. While shared vision-language representations provide strong synergy when video and text are semantically consistent (TV2A), they may amplify modality competition under semantic conflicts (TC-V2A), leading to the commonly observed visual dominance problem where textual control becomes ineffective.

To address these challenges, we propose a joint visual encoding paradigm that integrates two complementary visual encoders, enabling the model to capture both vision-language and audio-visual alignment. As illustrated in \cref{fig:controlfoley}, the input video is processed by a CLIP visual encoder and the proposed CAV-MAE-ST encoder simultaneously. The CLIP branch preserves strong vision-language semantic alignment, facilitating effective cooperation between visual and textual features when their semantics are consistent. In parallel, the CAV-MAE-ST branch produces audio-visual aligned representations that capture motion patterns and event-level cues closely related to sound generation.

The outputs of the two branches are fused to obtain the final visual representation:

\begin{equation}
z_v^{\mathrm{joint}} =
\mathrm{Proj}(z_v^{\mathrm{CLIP}}) +
\mathrm{Proj}(z_v^{\mathrm{CAV}}) .
\end{equation}

where $z_v^{\mathrm{CLIP}}$ and $z_v^{\mathrm{CAV}}$ denote the visual embeddings extracted from the CLIP encoder and the CAV-MAE-ST encoder, respectively.

This dual-branch design provides complementary advantages for different scenarios. When textual and visual semantics are consistent (TV2A), the CLIP branch promotes strong cross-modal synergy for high-quality generation. When semantic conflicts occur (TC-V2A), the CAV-MAE-ST branch introduces audio-visual aligned representations that mitigate excessive visual dominance and stabilize multimodal fusion, thereby improving textual controllability.

\subsection{Timbre-Focused Reference Audio Control}

In conventional V2A generation, visual inputs typically provide sufficient information to determine the temporal structure and basic acoustic events of the generated audio. However, in many practical scenarios, users may require additional control over the acoustic style of the generated sound, such as timbre characteristics or recording conditions. Similar to textual prompts that provide semantic guidance, reference audio offers an intuitive way to specify desired acoustic styles and enables more flexible and personalized audio generation.

To support such controllability, we extend the backbone with a timbre-focused reference audio control mechanism. The key idea is to utilize reference audio to convey acoustic style information while avoiding interference with the temporal structure determined by the input video. Specifically, our design introduces two complementary conditioning pathways: a reference audio conditioning branch within the multimodal transformer that provides token-level conditioning, and a global timbre conditioning signal that injects style-related information into the generation process.

\noindent\textbf{Global Semantic Conditioning.} To incorporate reference audio into the generation process, we introduce a dedicated conditioning branch within the multimodal transformer. The reference audio is first encoded by a pretrained CLAP encoder to obtain audio embeddings. These embeddings are projected into the shared latent space and injected into the MMDiT backbone through multimodal transformer blocks, following a design analogous to the textual conditioning pathway ~\cite{Cheng2025MMAudio}.

To ensure that the reference audio mainly provides acoustic style information rather than temporal dynamics, the branch is modified to suppress temporal modeling. Specifically, positional encodings for the reference audio tokens are removed, preventing the model from distinguishing temporal ordering within the sequence. In addition, modules originally designed to capture local temporal patterns are simplified by replacing ConvMLP blocks with standard MLP layers and using linear projections instead of temporal convolutions. These modifications reduce the model's ability to encode temporal structures from the reference audio and encourage the learned representation to focus on global acoustic characteristics.

\noindent\textbf{Global Timbre Conditioning.} In addition to token-level conditioning, we further introduce a global timbre conditioning signal. A pretrained audio StyleConditioner ~\cite{copet2023simple} encoder is used to extract timbre-related representations from the reference audio. The resulting embedding is projected into the latent space and injected as a global conditioning signal
$c_{\mathrm{timbre}} = \mathrm{Proj}(E_{\mathrm{timbre}}(a_r))$,
where $a_r$ denotes the reference audio, $E_{\mathrm{timbre}}(\cdot)$ represents the timbre encoder, and ${\mathrm{Proj}}(\cdot)$ is a learnable projection layer.

The global timbre embedding is combined with visual and textual conditioning signals to guide the generation process, enabling the synthesized audio to inherit acoustic style characteristics from the reference signal while maintaining temporal synchronization with the input video.

\subsection{Modality-Robust Training Strategies}

To improve robustness under diverse conditioning scenarios, the model should be able to operate with different combinations of available modalities. In practical applications, visual, textual, and reference audio inputs may not always be simultaneously provided. Therefore, the training process should encourage the model to remain stable and controllable even when some conditioning modalities are missing.

To achieve this goal, we adopt an all-modality random dropout strategy during training. Specifically, the visual, textual, and reference audio inputs are randomly dropped with predefined probabilities. This strategy enables the model to learn modality-agnostic representations and prevents over-reliance on any single modality, allowing the model to generalize to different conditioning configurations.

In addition to modality dropout, we further introduce a representation alignment objective based on REPA loss to improve multimodal semantic consistency. The goal is to encourage the intermediate audio representations within the diffusion transformer to remain semantically aligned with the conditioning signals.

A straightforward solution would be to align the generated audio representation with individual modality features (e.g., visual, textual, or reference audio representations). However, this strategy becomes problematic under modality dropout, where certain modality features may be absent during training.

To address this issue, we construct a unified multimodal alignment target by aggregating the available global visual, textual, and timbre representations. This aggregated representation naturally adapts to different modality combinations while avoiding modality-specific alignment logic.

Let $z_{\mathrm{audio}}$ denote the intermediate audio representation from the diffusion transformer and $z_{\mathrm{cond}}$ denote the aggregated multimodal feature obtained from visual, textual, and timbre representations. The REPA loss is defined as

\begin{equation}
\mathcal{L}_{\mathrm{REPA}} = -\frac{\mathrm{Proj}(h_{\text{audio}})\cdot z_{\text{cond}}}{\left |  \mathrm{Proj}(h_{\text{audio}})\right |\cdot  \left |z_{\text{cond}}\right |} 
\end{equation} 

where ${\mathrm{Proj}}(\cdot)$ is a learnable projection MLP.

Notably, the aggregated multimodal representation is derived from the same global conditioning features already used by the diffusion transformer. Therefore, the proposed alignment objective introduces no additional modality encoder and fully exploits the existing multimodal representations within the model. This design simplifies the training pipeline while encouraging the generated audio features to remain semantically consistent with the available conditioning signals under diverse modality configurations.

\subsection{VGGSound-TVC: A Benchmark for Assessing Text Controllability under Visual-Text Semantic Conflict}

\begin{table*}[t]
\caption{Example samples from the VGGSound-TVC benchmark. Each video is associated with four textual descriptions exhibiting different levels of visual-text semantic conflict.}
\label{tab:vggsound_tvc_examples}
\centering
\small
\begin{tabular}{c|c|c|c|c|c}
\hline
\textbf{video\_id} & \textbf{L0} & \textbf{L1-subject} & \textbf{L1-action} & \textbf{L2} & \textbf{L3}\\
\hline

zpWuikVorYg\_000032 &
elk bugling  &
wolf howling  &
elk screaming  &
air raid siren  &
dog barking  \\

RgyqhpOJFM4\_000030 &
playing cello  &
playing violin  &
scratching cello  &
sawing wood  &
people battle cry  \\

UyCw7pCgYg8\_000055 &
tap dancing  &
horse trotting  &
typing on keyboard  &
rain on tin roof  &
bathroom ventilation fan running  \\

\hline
\end{tabular}
\end{table*}

Current TV2A methods typically adopt a bimodal conditional generation paradigm, where models receive both video inputs and textual descriptions as generation conditions. In this framework, the textual modality should serve as an important control signal that guides the semantic content and event categories of the generated audio.

However, existing TV2A training and evaluation datasets are predominantly constructed under a video-text semantic alignment assumption, where the textual description closely matches the visual content ~\cite{Chen2020VGGSound, Polyak2024MovieGen, shan2025hunyuanvideofoley, liu2025thinksound, tian2026audiox, jeong2025rewas,xing2024seeandhear, liu2024vatt}. In such scenarios, the visual and textual modalities often provide redundant information. As a result, models may still produce plausible audio outputs by relying primarily on visual cues, even when the textual condition is weakened or ignored. This characteristic makes it difficult to properly evaluate the degree to which models rely on textual inputs, and limits our ability to analyze modality dominance when visual and textual signals conflict.

To address this limitation, we introduce VGGSound-TVC (Text-Visual Conflict), a benchmark designed for assessing text controllability under visual-text semantic conflict. The benchmark is constructed based on the VGG-SS dataset, 
a manually curated subset of the VGGSound‑Test set with no overlap against the training set. Designed for video‑level audio‑visual sound source localization, it contains video frames in which sounding objects are clearly visible ~\cite{chen2021vggss}. Building upon this dataset, we systematically modify textual descriptions to introduce varying levels of semantic conflict with the visual content. This design intentionally breaks the original video-text alignment and forces models to balance visual and textual information when generating audio.

Furthermore, we define four levels of visual-text semantic conflict, ranging from no conflict (L0) to strong conflict (L3). By gradually increasing the conflict strength, the VGGSound-TVC benchmark enables systematic analysis of modality reliance under different levels of multimodal inconsistency, and allows us to quantify how text controllability changes as the degree of visual-text conflict increases.

\noindent\textbf{Dataset Description.} The proposed benchmark contains a total of 25,005 video-text pairs. It is constructed from 5,001 videos, where each video is associated with five textual variants (four conflict levels L0--L3, where L1 includes two variants),
resulting in $5{,}001 \times 5 = 25{,}005$ samples.

Each sample consists of a video and its corresponding textual label. The dataset provides the following textual fields:

\begin{itemize}
\item \textbf{label\_L0}: A normalized textual label in the format of ``subject + action'', obtained by simplifying and standardizing the original multi-label annotations.

\item \textbf{label\_L1\_subject}: A mild semantic conflict introduced at the subject level, where the action description remains unchanged while the sounding subject is replaced.

\item \textbf{label\_L1\_action}: A mild semantic conflict introduced at the action level, where the subject remains unchanged while the action description is modified.

\item \textbf{label\_L2}: A moderate semantic conflict in which the textual description belongs to a different semantic category while still maintaining a similar temporal structure or acoustic rhythm.

\item \textbf{label\_L3}: A strong conflict generated by randomly replacing the category label while preserving the overall class distribution.
\end{itemize}

The L0, L1, and L2 textual labels are generated using the Gemini 2.5 Pro multimodal large language model ~\cite{comanici2025gemini}, followed by rule-based filtering and normalization. The detailed prompt we meticulously designed and input to Gemini 2.5 Pro for generating these textual variants is provided in \cref{appendix:propmpt}. To verify the accuracy of labels generated by Gemini 2.5 Pro, human validation is conducted with five annotators on 100 source videos covering all five conflict levels (2,500 ratings). Intended conflict levels match human judgments with 94.6\% agreement; pairwise agreement is 90.2\% (PABAK=0.80); text naturalness is 4.94/5. These results support VGGSound-TVC's conflict-level validity and prompt naturalness.

\noindent\textbf{Example Samples.} \cref{tab:vggsound_tvc_examples} presents several example samples from the VGGSound-TVC benchmark, illustrating different levels of visual-text semantic conflict.

\begin{table*}[t]
\centering
\caption{Data construction statistics of AC‑VAS dataset.}
\label{tab:av_vas}
\setlength{\tabcolsep}{4pt}
\begin{tabular}{l|c|c|c|c|c}
\hline
Subset
& \makecell{Pairing Strategy}
& \#Target Classes
& \#Targets per Class
& \#References per Target
& Total Pairs \\
\hline
same-class
& \makecell[l]{Reference belongs to the same class\\as target (no overlap)}
& 8 & 10 & 3 & $8\times10\times3=240$ \\
\hline
cross-near
& \makecell[l]{References sampled from near‑neighbor\\classes within the same acoustic family\\(different class)}
& 8 & 10 & 3 & $8\times10\times3=240$ \\
\hline
cross-far
& \makecell[l]{References sampled from far classes\\of different acoustic families}
& 8 & 10 & 3 & $8\times10\times3=240$ \\
\hline
\end{tabular}
\end{table*}

\section{Experiments}

\subsection{Experimental Settings}

Experimental settings consist of three components: evaluation datasets, evaluation metrics, and baselines for comparison, which are introduced one by one below.

\noindent\textbf{Datasets.} We evaluated our method using multiple benchmarks in three different tasks: TV2A, TC-V2A, and AC-V2A.

For the TV2A task, we adopt three evaluation datasets: VGGSound-Test ~\cite{Chen2020VGGSound}, Kling-Audio-Eval ~\cite{wang2025klingfoley}, and MovieGen-Audio-Bench \footnote{This project uses the MovieGen-Audio-Bench dataset (\href{https://github.com/facebookresearch/MovieGenBench}{MovieGenBench}), which is licensed under the \href{https://creativecommons.org/licenses/by-nc/4.0/}{CC BY-NC 4.0}. The authors confirm that the use of the aforementioned dataset in this project is used for academic and non-commercial purposes only.} ~\cite{Polyak2024MovieGen}.
Their characteristics are as follows: VGGSound-Test shares the same origin as our training set VGGSound-Train, both coming from the VGGSound dataset, and serves as an in-distribution benchmark. In contrast, Kling-Audio-Eval, which features fine-grained audio annotations and rich category coverage, and MovieGen-Audio-Bench , which provides cinematic multi-scene content, offer complementary out-of-distribution evaluation scenarios. These three datasets complement each other in terms of data distribution, scene complexity, and annotation granularity, enabling a comprehensive and systematic validation of the model’s generation quality and generalization ability under diverse conditions. 

For the TC-V2A task, the VGGSound-TVC dataset presented in Chapter 3 is employed to evaluate the text controllability of the model. 

For the AC-V2A task, we first utilize the public Greatest Hits dataset~~\cite{owens2016greatesthits}. This dataset consists of recordings of drumsticks striking objects made of different materials and possesses two key characteristics: first, high temporal synchronization between audio and video, ensuring precise alignment between visual actions and audio events; second, rich audio texture, encompassing the distinct timbres produced by various materials such as wood, metal, and plastic. This provides ideal data for evaluating a model’s audio-conditional generation capabilities. We adopted the methods from CondFoleyGen ~\cite{du2023condfoleygen}, selecting 194 silent input videos from the 977 videos in the Greatest Hits dataset and pairing each with three conditional audio tracks from different test videos. This yielded a conditional sound effects evaluation dataset comprising 582 input video-conditional audio pairs, covering 17 different materials and two action types. Each silent video and conditional audio clip is 2 seconds in length. We further divide these pairs based on whether the actions are identical (material‑only conflict) or different (material‑plus‑action conflict). 
Next, to evaluate the model’s performance under diverse sound events, we construct AC‑VAS, a 720-pair evaluation set from visually aligned VAS videos ~~\cite{chen2020generating}, covering eight broader classes (baby/cough/dog/drum/fireworks/gun/hammer/sneeze) from vocal/animal, impact, and explosive acoustic families. The vocal/animal family group (baby, cough, sneeze, dog) exhibits harmonics, formants, airflow bursts or barking. The impact family group (drum, hammer) has short transients and decay envelopes. The explosive family group (fireworks, gun) features extremely short attack and broadband noise tails. We construct the AC‑VAS dataset following these acoustic‑family groupings and further partition it into three subsets. The silent input video is denoted as the target, and the reference audio as the reference. The detailed data‑construction protocol is presented in \cref{tab:av_vas}. In particular, the same‑class, cross‑near, and cross‑far subsets share the same 80 target videos, facilitating direct cross‑subset comparison.

\begin{figure*}[t]
  \centering
  \includegraphics[width=0.9\linewidth]{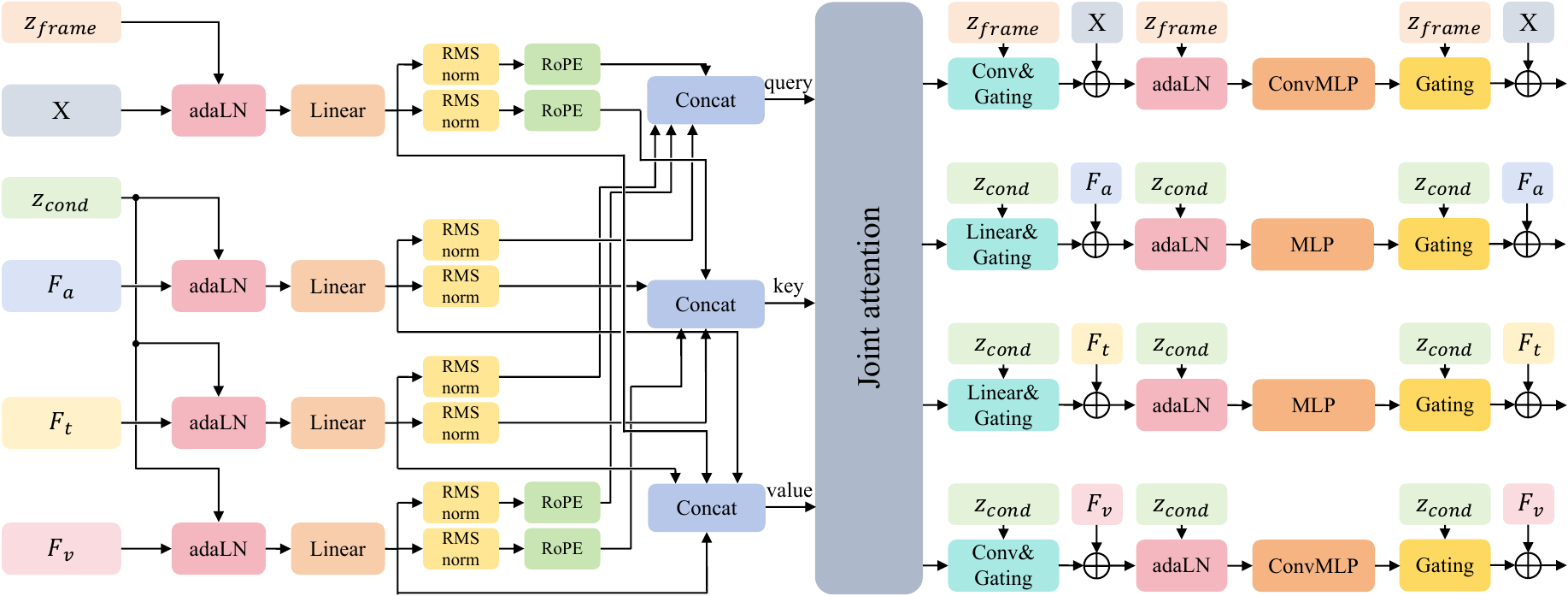}
  \caption{The internal architecture of the Multimodal Transformer block.}
  \label{fig:MMDiT}
\end{figure*}

\noindent\textbf{Evaluation Metrics.} Similarly, we have designed different evaluation metrics for the three types of tasks.

For the TV2A task, we conduct a comprehensive evaluation based on the following aspects: semantic alignment between the generated audio and the input video and text content; temporal synchronization between the generated audio and the input visual frames; distribution matching between the generated audio and real audio; and the quality of the generated audio itself. The evaluation metrics for semantic alignment are the IB-score and the CLAP-score. The former is calculated as the cosine similarity between the visual features of the input video and the audio features of the generated audio, both extracted by the ImageBind model ~\cite{girdhar2023imagebind}; a higher score indicates greater semantic alignment between the generated audio and the input video. The latter is derived from the cosine similarity between the text features extracted from the input text description by the CLAP model and the audio features of the generated audio; similarly, a higher score indicates stronger semantic alignment between the generated audio and the input text description. We employ two CLAP models---LAION-CLAP ~\cite{wu2023laionclap} and MS-CLAP ~\cite{elizalde2023msclap}---to extract these features. For temporal synchronization evaluation, we adopt the DeSync metric. Specifically, we use the Synchformer ~\cite{iashin2024synchformer} model to extract features from the generated audio and input visual content, and then measure the temporal offset between them. A smaller temporal offset indicates better synchronization performance. Distribution matching is used to measure the distribution similarity between generated audio and real audio. We use the PANNs ~\cite{kong2020panns} and PaSST ~\cite{koutini2021passt} models to extract audio features, and calculate the Kullback-Leibler divergence between the generated audio and the target audio as the evaluation metric. Specifically, PANNs operates at 16 kHz, while PaSST operates at 32 kHz. In addition, The audio quality metric IS is independent of real audio-visual data. It evaluates generated audio based on the prediction confidence of category classifiers and the diversity of category distributions. A higher score indicates better audio quality, i.e., clearer and more diverse samples. We adopt the PANNs classifier and the PaSST classifier to conduct this evaluation.

For the TC-V2A task, we evaluate the text controllability under conflicting visual and text inputs by analyzing the trends of IB and CLAP scores across different conflict levels, as well as the DeSync metric for temporal consistency. The definitions and calculation methods of these metrics are consistent with those described above.

For the AC-V2A task, we aim to evaluate the timbre consistency between the generated audio and the input reference audio, as well as whether the generated audio aligns with the actions in the input video. To this end, we employ Resemblyzer ~\cite{wan2018resemblyzer} to extract features from the generated and reference audios, and calculate their cosine similarity to quantify timbre similarity together with time‑pooled MFCC and PaSST‑based cosine similarity (CosSim). Meanwhile, we adopt the DeSync to evaluate audio-visual synchronization.

\noindent\textbf{Baselines.} To comprehensively verify the effectiveness of the proposed method, we select five representative baseline models covering different technical routes for audio generation from the state-of-the-art (SOTA) models of TV2A, TC-V2A and AC-V2A tasks. The core characteristics and application scenarios of each model are elaborated as follows:

As a classic multimodal audio generation model, MMAudio ~\cite{Cheng2025MMAudio} is capable of generating high-fidelity, frame-synchronized audio based on video or text. It has laid the groundwork for subsequent models and remains state-of-the-art across multiple metrics. Therefore, we selected the best-performance “Large” version of this model as the baseline for the TV2A and TC-V2A tasks.

ThinkSound ~\cite{liu2025thinksound} and HunyuanVideo-Foley ~\cite{shan2025hunyuanvideofoley}, as recently proposed models tailored for the V2A task, are applicable to both TV2A and TC-V2A tasks. The former supports step-by-step interactive audio generation, while the latter is optimized to address conflicts between visual and text modalities. It is worth noting that, ThinkSound, since the official release does not provide a pre-trained large language model component for generating Chain-of-Thought (CoT) instructions, we evaluated only the version that does not include CoT instructions. For HunyuanVideo-Foley, we choose the best "XXL" version for comparison.

As the only baseline model applicable to all three tasks in our experiments, AudioX ~\cite{tian2026audiox} supports the arbitrary combination of text, image, and audio as inputs and enables the generation from these modalities to audio. The key advantage of this model lies in its strong generalization ability across different tasks. We choose the best "MAF-MMDiT" version of AudioX for comparison. Thus, we can effectively validate the overall performance of our proposed method in diverse task scenarios.

Besides the above general‑purpose models, to better compare the performance of AC‑V2A, this paper introduces a dedicated model CondFoleyGen ~\cite{du2023condfoleygen}. Given a pair of audio‑visual inputs, this model generates audio that preserves the timbre characteristics of the reference audio and is precisely synchronized with the motion in the input video. Therefore, CondFoleyGen is selected as the dedicated baseline for the AC-V2A task, providing a direct comparison standard for evaluating the timbre consistency and audio-visual alignment performance of our method in audio-conditional generation scenarios. MultiFoley ~\cite{Chen2025Multifoley} provides similar functionality, however, it is not open-source, so it is not included in our comparative experiments.

All baseline models are directly adopted from their officially released pre-trained weights. When multiple pre-trained versions of a model are available, we consistently select the optimal performing version for the benchmark comparisons. All baseline models and the proposed method in this paper perform inference and evaluation under exactly the same hardware environment.

\subsection{Training Details}

\begin{table*}[t]
\caption{Quantitative results on the TV2A task across VGGSound-Test, Kling-Audio-Eval, and MovieGen-Audio-Bench. The best and second-best results are highlighted in bold and underlined, respectively.}
\centering
\renewcommand{\arraystretch}{1.0}
\setlength{\tabcolsep}{2.4pt}

\begin{tabular}{p{1.7cm}p{4.1cm}cccccccc}
\toprule
Dataset & Method 
& \multicolumn{3}{c}{Semantic alignment} 
& Temporal sync
& \multicolumn{2}{c}{Audio quality} 
& \multicolumn{2}{c}{Distribution matching} \\

\cmidrule(lr){3-5}
\cmidrule(lr){6-6}
\cmidrule(lr){7-8}
\cmidrule(lr){9-10}

& 
& IB $\uparrow$
& CLAP$_{\text{LAION}} \uparrow$
& CLAP$_{\text{MS}} \uparrow$
& DeSync $\downarrow$
& IS$_{\text{PANNs}} \uparrow$
& IS$_{\text{PaSST}} \uparrow$
& KL$_{\text{PANNs}} \downarrow$ 
& KL$_{\text{PaSST}} \downarrow$ \\

\midrule

\multirow{5}{*}{\makecell{VGGSound\\Test}}

& MMAudio-L~~\cite{Cheng2025MMAudio}
& \textbf{0.33} & 0.22 & \underline{0.31}
& \underline{0.45}
& \underline{17.36} & \underline{13.31}
& \textbf{1.66} & \textbf{1.42} \\

& HunyuanVideo-Foley-XXL~~\cite{shan2025hunyuanvideofoley}
& \underline{0.32} & \underline{0.23} & 0.28
& 0.55
& 15.26 & 11.30
& 2.02 & 1.77 \\

& ThinkSound (w/o CoT)~~\cite{liu2025thinksound}
& 0.24 & 0.17 & 0.27
& 0.56
& 11.61 & 8.52
& 1.83 & 1.62 \\

& AudioX-MAF-MMDiT~~\cite{tian2026audiox}
& 0.28 & 0.19 & 0.28
& 0.89
& 15.83 & 11.44
& 2.02 & 1.77 \\

& \textbf{ControlFoley (ours)}
& \underline{0.32} & \textbf{0.26} & \textbf{0.36}
& \textbf{0.42}
& \textbf{22.08} & \textbf{15.87}
& \underline{1.71} & \underline{1.43} \\

\midrule

\multirow{5}{*}{\makecell{Kling\\Audio-Eval}}

& MMAudio-L~~\cite{Cheng2025MMAudio}
& \underline{0.30} & 0.20 & \underline{0.37}
& \underline{0.55}
& \underline{8.54} & \underline{7.03}
& 2.41 & \underline{2.03} \\

& HunyuanVideo-Foley-XXL~~\cite{shan2025hunyuanvideofoley}
& \textbf{0.34} & \underline{0.23} & 0.36
& \underline{0.55}
& 8.42 & 6.85
& \textbf{2.00} & \textbf{1.63} \\

& ThinkSound (w/o CoT)~~\cite{liu2025thinksound}
& 0.23 & 0.20 & 0.35
& 0.72
& 6.97 & 5.70
& 2.44 & 2.15 \\

& AudioX-MAF-MMDiT~~\cite{tian2026audiox}
& 0.26 & 0.20 & 0.33
& 0.93
& 8.89 & 6.73
& 2.50 & 2.13 \\

& \textbf{ControlFoley (ours)}
& 0.28 & \textbf{0.24} & \textbf{0.38}
& \textbf{0.52}
& \textbf{9.09} & \textbf{7.91}
& \underline{2.32} & 2.07 \\

\midrule

\multirow{5}{*}{\makecell{MovieGen\\Audio-Bench}}

& MMAudio-L~~\cite{Cheng2025MMAudio}
& 0.26 & 0.29 & 0.43
& 0.75
& \textbf{8.59} & 6.48
& 2.99 & 2.53 \\

& HunyuanVideo-Foley-XXL~~\cite{shan2025hunyuanvideofoley}
& \textbf{0.31} & \underline{0.31} & \underline{0.45}
& \underline{0.70}
& \underline{8.50} & \underline{6.50}
& \textbf{2.63} & \textbf{2.18} \\

& ThinkSound (w/o CoT)~~\cite{liu2025thinksound}
& 0.18 & 0.19 & 0.36
& 0.84
& 5.41 & 4.36
& 3.28 & 3.04 \\

& AudioX-MAF-MMDiT~~\cite{tian2026audiox}
& 0.23 & 0.28 & 0.42
& 1.06
& 7.81 & 5.74
& 2.81 & 2.46 \\

& \textbf{ControlFoley (ours)}
& \underline{0.27} & \textbf{0.34} & \textbf{0.46}
& \textbf{0.69}
& \underline{8.50} & \textbf{7.22}
& \underline{2.73} & \underline{2.38} \\

\bottomrule
\end{tabular}

\label{tab:v2a_results}
\end{table*}

We trained two models: the audio-visual encoder CAV-MAE-ST and the unified multimodal V2A model ControlFoley. 

\noindent\textbf{CAV-MAE-ST Training.} We train the CAV-MAE-ST encoder using the full training split of the VGGSound ~\cite{Chen2020VGGSound} dataset, which contains 180K 10-second video clips across 309 categories, with each clip associated with semantically consistent audio-visual pairs. Specifically, during training, videos are sampled at 4 fps, and Mel-filterbank features are extracted with 128 mel bins and a 10 ms frame shift. For each video frame, we crop an audio segment of approximately 2 seconds (corresponding to 208 time frames) centered at its corresponding timestamp in the Mel spectrogram, forming fine-grained audio-visual pairs. At the model level, both the unimodal encoders and the multimodal encoder-decoder adopt the Vision Transformer (ViT) architecture ~\cite{dosovitskiy2021vit}, with 75\% of tokens randomly masked for each modality. For the loss settings, we set the weights of the contrastive loss and the reconstruction loss to 0.01 and 1, respectively. Training is performed on GPU devices with a total computing power of 176 TFLOPS (FP32 precision), using a learning rate of 1e-4 and a batch size of 160 until convergence, requiring approximately 150 epochs.

\noindent\textbf{ControlFoley Training.} The training data comprises audio-visual-text and audio-text data in a 1:1 ratio. The audio-visual-text data corresponds to the aforementioned VGGSound training set, which is repeated to 5 times its original size. The audio-text data includes AudioCaps ~\cite{kim2019audiocaps}, WavCaps ~\cite{mei2024wavcaps}, and Clotho ~\cite{drossos2020clotho}, amounting to 900K clips in total. Audio, visual, and textual features are extracted from these datasets. For modalities lacking in the datasets, we represent their features using learnable null tokens. Additionally, we randomly mask modality inputs with a probability of 10\% to improve the robustness of the model. 

The construction strategy for reference audio during training is detailed below.
Specifically, the ground-truth target audio serves as the reference audio. For the reference audio conditioning branch, the full audio is encoded using a pre-trained CLAP encoder, yielding features with no temporal dimension. For the global timbre conditioning, a random 2--4 second clip is sampled from the reference audio, then encoded by a pre-trained audio StyleConditioner encoder.

Our model backbone consists of 18 multimodal DiT blocks and 36 unimodal DiT blocks. \cref{fig:MMDiT} illustrates the internal architecture of the Multimodal Transformer block in our proposed model ControlFoley, where $X$ denotes the audio latents, $F_{v}$ and $F_{t}$ represent visual features and textual features, $F_{a}$ denotes the reference audio feature, $z_{\mathrm{cond}}$ stands for the multimodal global condition, and $z_{\mathrm{frame}}$ refers to the frame-aligned synchronization condition. The queries, keys, and values from the four branches are concatenated sequentially for joint attention, and the output is split back into four paths according to the input order. 
Furthermore, during the pre-attention and post-attention stages, we perform adaptive LayerNorm, RMSNorm, and other operations on each branch, while selectively applying RoPE and other temporal-related operations according to the temporal correlation of each branch. 

We use the hidden states output by the 8th unimodal DiT block for representation alignment, aiming to strike a balance between preserving fine-grained audio details and achieving semantic alignment. Similarly, this generative model is trained for 300,000 iterations on GPU devices with a total computing power of 176 TFLOPS (FP32 precision). During inference, we adopt 25 inference steps and a classifier-free guidance scale of 4.5 to generate audio with a sampling rate of 44.1 kHz.

\subsection{Main Results}
\subsubsection{TV2A}

\begin{figure}[t]
  \centering
  \includegraphics[width=\linewidth]{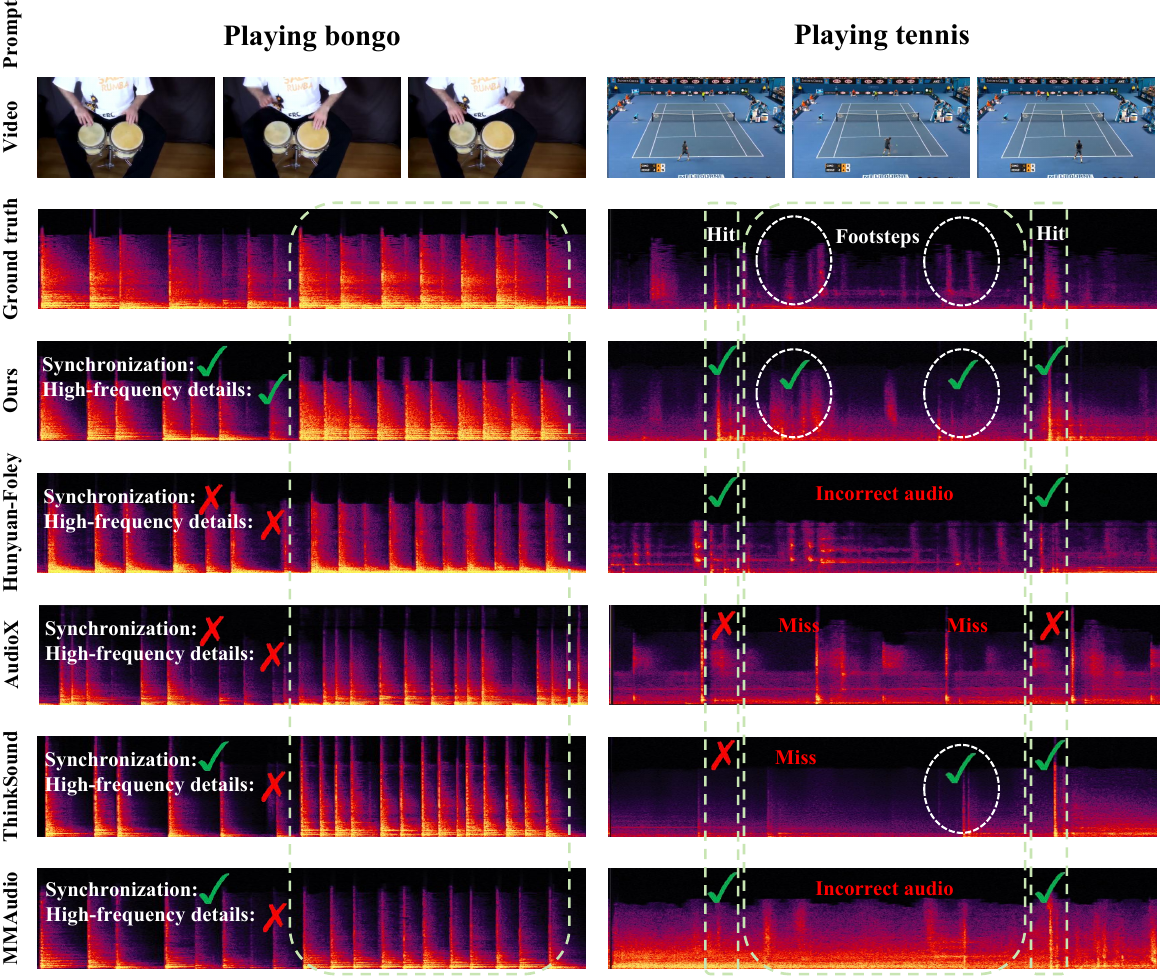}
  \caption{Spectrogram comparison on the TV2A task. 
Left: “Playing the bongo”, characterized by dense and high-frequency patterns. 
Right: “Playing tennis”, involving multiple sound events (e.g., hits and footsteps). 
Our method achieves better temporal synchronization and recovers richer high-frequency details than prior methods.}
  \label{fig:tv2a}
\end{figure}

We evaluate ControlFoley on the text-guided video-to-audio (TV2A) task across three benchmarks: VGGSound-Test, Kling-Audio-Eval, and MovieGen-Audio-Bench, covering both in-distribution and out-of-distribution scenarios. The quantitative results are presented in \cref{tab:v2a_results}. Representative spectrogram comparisons are shown in \cref{fig:tv2a}.

\noindent\textbf{Overall Performance.}
ControlFoley achieves state-of-the-art performance across all benchmarks, delivering the best audio-text semantic alignment, temporal synchronization, and audio quality. In particular, it consistently attains the highest CLAP scores on all datasets, while also achieving the lowest DeSync. Moreover, ControlFoley significantly improves audio quality, achieving up to 27\% relative gain in IS (22.08 vs.\ 17.36 on VGGSound), with consistent gains across all datasets.

\noindent\textbf{Semantic Alignment.}
As shown in \cref{tab:v2a_results}, ControlFoley consistently ranks first or second across all metrics. While some methods obtain slightly higher IB values, their lower CLAP scores suggest an over-reliance on visual information. In contrast, ControlFoley achieves a better balance by selectively preserving relevant visual cues while prioritizing text-aligned audio generation.

\noindent\textbf{Training Efficiency.}
Despite its strong performance across benchmarks, ControlFoley is trained on a relatively small-scale dataset, consisting of approximately 180K audio-visual-text pairs and 900K audio-text pairs, all from publicly available sources. In contrast, HunyuanVideo-Foley is trained on around 100M proprietary audio-visual-text samples, while AudioX leverages approximately 7M audio-visual-text pairs. Meanwhile, methods such as MMAudio and ThinkSound are trained on datasets of a similar scale to ours (approximately 180K--300K audio-visual-text pairs and 700K--900K audio-text pairs).

These results highlight that ControlFoley not only benefits from strong data efficiency compared to models trained on substantially larger datasets, but also achieves superior performance under comparable training data settings, demonstrating robust generalization ability.

\begin{figure}[t]
\centering
\includegraphics[width=\linewidth,height=0.15\paperheight]{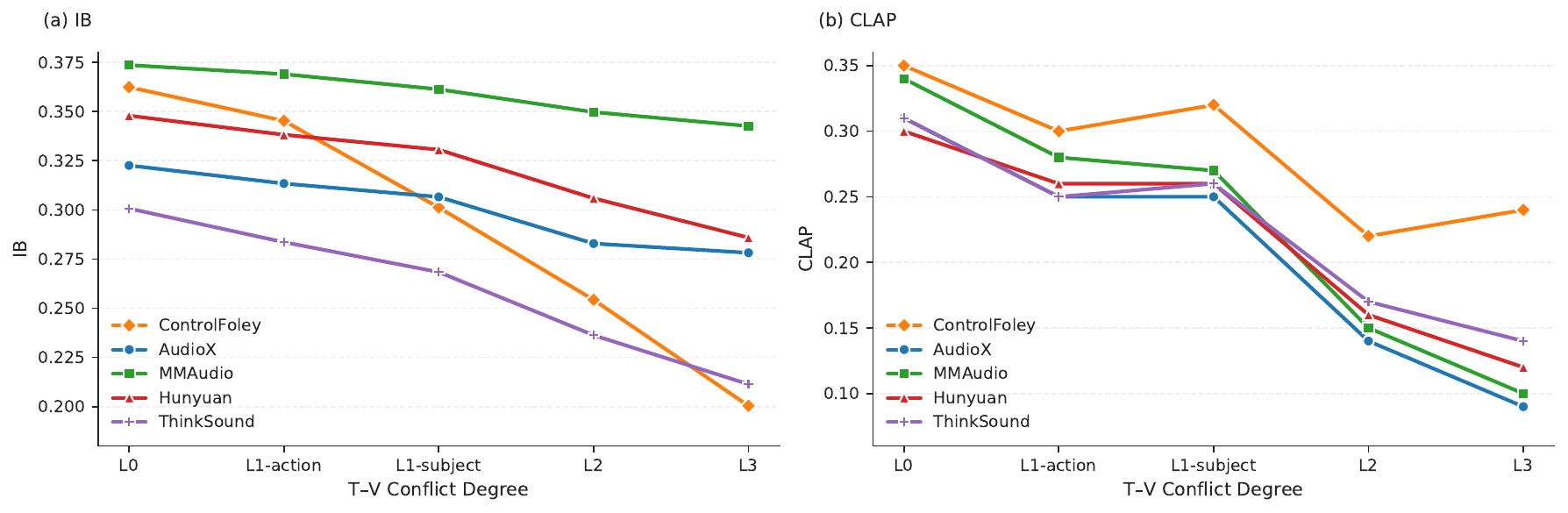}
\caption{Performance under different T-V Conflict Degrees.
(a) IB. (b) CLAP. At L0 (no conflict), higher IB and CLAP indicate better semantic alignment with video and text. As the conflict increases, lower IB reflects stronger deviation from the generated audio, indicating better text controllability, while higher CLAP remains preferable.}
\label{fig:tc_v2a}
\end{figure}

\begin{table}[t]
\caption{Audio-video temporal synchronization (DeSync $\downarrow$) under different conflict levels (lower is better). Best results are in bold.}
\centering
\setlength{\tabcolsep}{2pt}

\begin{tabular}{lccccc}
\toprule
Method & L0 & L1-act & L1-sub & L2 & L3\\
\midrule
AudioX-MAF-MMDiT~~\cite{tian2026audiox} & 0.86 & 0.86 & 0.86 & 0.87 & 0.88\\
MMAudio-L~~\cite{Cheng2025MMAudio} & 0.39 & 0.38 & 0.39 & \textbf{0.38} & \textbf{0.39}\\
HunyuanVideo-Foley-XXL~~\cite{shan2025hunyuanvideofoley} & 0.51 & 0.51 & 0.52 & 0.52 & 0.52\\
ThinkSound~~\cite{liu2025thinksound} & 0.50 & 0.48 & 0.48 & 0.50 & 0.51\\
\textbf{ControlFoley (ours)} & \textbf{0.37} & \textbf{0.36} & \textbf{0.37} & \textbf{0.38} &  0.41\\
\bottomrule
\end{tabular}
\label{tab:tcv2a_tempsynch}
\end{table}

\subsubsection{TC-V2A}

\begin{figure}[t]
  \centering
  \includegraphics[width=0.9\linewidth]{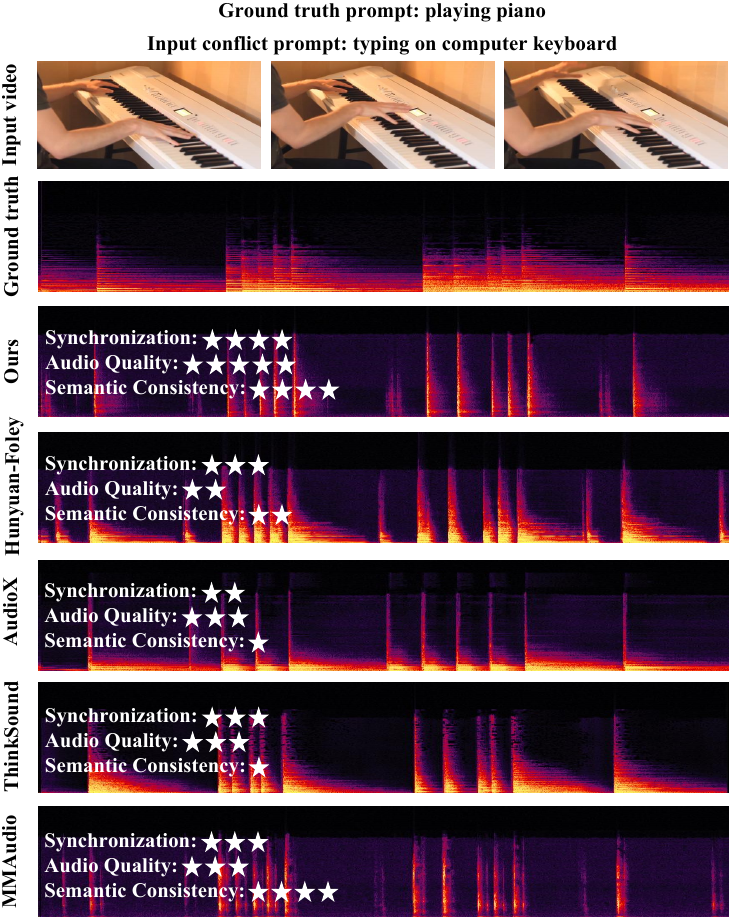}
  \caption{Comparison on the TC-V2A task under conflicting video-text inputs 
(video: “playing the piano”, text: “typing on keyboard”).  
ControlFoley better preserves text-consistent acoustic patterns while maintaining temporal alignment.}
  \label{fig:tcv2a}
\end{figure}

We further evaluate text-controllable video-to-audio generation (TC-V2A) under different text-visual (T-V) conflict levels, where the textual description progressively deviates from the visual content. This setting explicitly tests whether a model can adaptively adjust modality reliance when conflicts arise.
Quantitative results are shown in \cref{fig:tc_v2a} and \cref{tab:tcv2a_tempsynch}, with representative cases in \cref{fig:tcv2a}.

\noindent\textbf{Modality Adaptation Behavior.}
We first analyze how models respond to increasing cross-modal conflict using IB and CLAP, as shown in \cref{fig:tc_v2a}. At L0 (no conflict), higher IB and CLAP indicate better alignment with both visual and textual inputs. As the conflict increases, a desirable model should reduce reliance on visual information (reflected by decreasing IB) while maintaining strong text alignment (reflected by consistently high CLAP).

ControlFoley clearly exhibits this desirable behavior. Specifically, IB decreases more rapidly than all baselines as the conflict level increases, indicating that the model effectively suppresses conflicting visual cues. Meanwhile, ControlFoley maintains consistently higher CLAP scores, especially under moderate conflict (L1), demonstrating stronger text controllability. In contrast, baseline methods tend to either over-preserve visual information (slow IB reduction) or suffer from degraded semantic alignment (lower CLAP), indicating limited ability to adaptively balance modalities.

\noindent\textbf{Temporal Synchronization under Conflict.}
As shown in \cref{tab:tcv2a_tempsynch}, ControlFoley achieves the best DeSync performance across almost all conflict levels. MMAudio has slightly lower DeSync under high conflict levels, but poor CLAP/high IB show text has little effect and visual dominance preserves synchronization. This indicates that ControlFoley maintains strong temporal alignment even when adapting to conflicting textual guidance.

\noindent\textbf{Summary.}
These results demonstrate that ControlFoley not only achieves superior text controllability under cross-modal conflict, but also preserves accurate temporal synchronization, effectively balancing modality adaptation and generation quality.

\subsubsection{AC-V2A}

\begin{figure}[t]
  \centering
  \includegraphics[width=\linewidth]{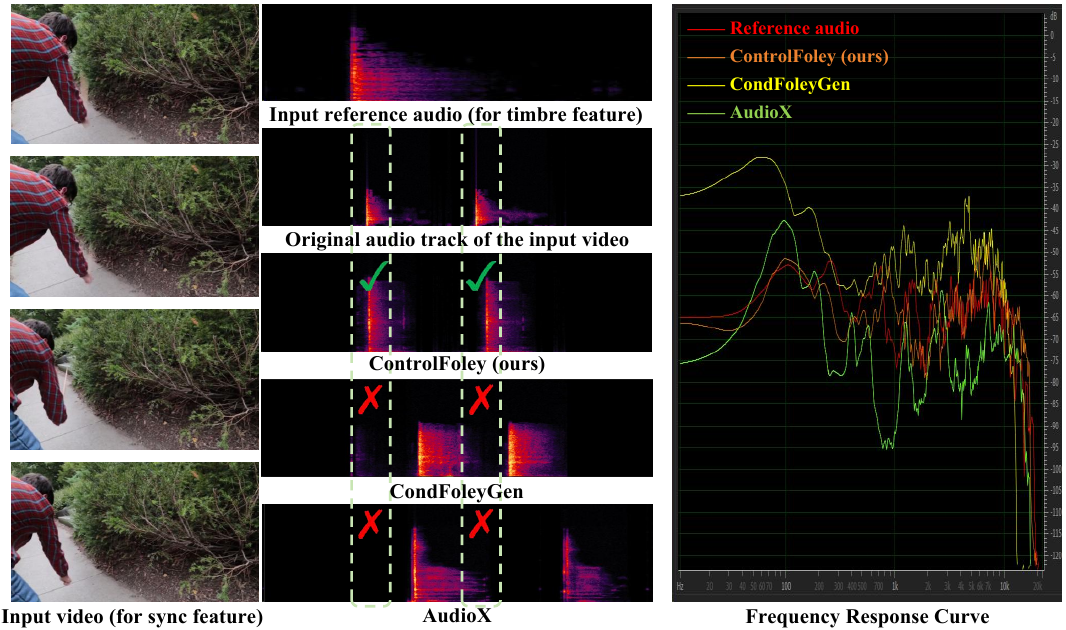}
  \caption{Spectrogram and frequency response comparison on the AC-V2A task. 
Left: input video. Middle: generated audio vs. original audio. Right: frequency response curves. 
Our method achieves more accurate temporal alignment with the video content and better matches the timbre of the reference audio.}
  \label{fig:acv2a}
\end{figure}

We further evaluate ControlFoley on the audio-controlled video-to-audio (AC-V2A) task using the Greatest Hits ~\cite{owens2016greatesthits} dataset. Following prior work, we compare with CondFoleyGen, a task-specific AC-V2A model trained on the same dataset, and AudioX as a general-purpose baseline. Representative results are shown in \cref{fig:acv2a}.

\noindent\textbf{Quantitative Results.}
As shown in \cref{tab:acv2a_results}, ControlFoley achieves the best performance on all evaluation metrics across both datasets. Notably, ControlFoley outperforms CondFoleyGen, a task‑specific model trained on data from the same source as the Greatest Hits test set. The advantage is particularly prominent on the out‑of‑domain AC‑VAS dataset with broader categories, demonstrating stronger generalization compared with task‑tailored designs. Compared with AudioX, which adopts a generic multimodal fusion strategy, ControlFoley achieves better timbre similarity and synchronization, suggesting that explicit disentanglement is crucial for precise audio-conditioned generation. 

\begin{table}[t]
\caption{Quantitative comparison on the AC‑V2A task over Greatest Hits and AC‑VAS datasets. Best results are in bold.}
\centering
\setlength{\tabcolsep}{2.0pt}
\begin{tabular}{llcccc}
\toprule
Conflict Split & Method & DeSync$\downarrow$ & Res.$\uparrow$ & MFCC$\uparrow$ & CosSim$\uparrow$\\
\midrule
\multicolumn{6}{l}{\textit{Greatest Hits dataset}} \\
\midrule
\textit{Material} & CondFoleyGen & 0.93 & 0.80 & 0.99 & \textbf{0.76}\\
only & AudioX & 0.97 & 0.70 & 0.99 & 0.57\\
 & \textbf{ControlFoley} & \textbf{0.84} & \textbf{0.81} & \textbf{1.00} & 0.74\\
\cmidrule{1-6}
\textit{Material+} & CondFoleyGen & 0.99 & 0.77 & 0.99 & 0.72\\
\textit{Action} & AudioX & 0.92 & 0.69 & 0.98 & 0.53\\
 & \textbf{ControlFoley} & \textbf{0.85} & \textbf{0.80} & \textbf{0.99} & \textbf{0.72}\\
\midrule
\multicolumn{6}{l}{\textit{AC‑VAS dataset}} \\
\midrule
Same & CondFoleyGen & 1.26 & 0.63 & 0.66 & 0.23\\
class & AudioX & 1.29 & 0.81 & 0.82 & 0.42\\
 & \textbf{ControlFoley} & \textbf{0.31} & \textbf{0.83} & \textbf{0.89} & \textbf{0.65}\\
\cmidrule{1-6}
Cross & CondFoleyGen & 1.29 & 0.63 & 0.68 & 0.24\\
near & AudioX & 1.28 & \textbf{0.82} & 0.85 & 0.42\\
 & \textbf{ControlFoley} & \textbf{0.37} & \textbf{0.82} & \textbf{0.91} & \textbf{0.57}\\
\cmidrule{1-6}
Cross & CondFoleyGen & 1.24 & 0.62 & 0.56 & 0.21\\
far & AudioX & 1.29 & \textbf{0.81} & 0.79 & 0.39\\
 & \textbf{ControlFoley} & \textbf{0.44} & \textbf{0.81} & \textbf{0.86} & \textbf{0.52}\\
\bottomrule
\end{tabular}
\label{tab:acv2a_results}
\end{table}

\noindent\textbf{Timbre Consistency.}
ControlFoley achieves the highest timbre similarity, indicating its ability to faithfully preserve audio characteristics under conditioning, while simultaneously maintaining accurate synchronization and high perceptual quality.

These results demonstrate that ControlFoley supports high-quality audio-conditioned generation, achieving strong timbre consistency without compromising synchronization or audio quality.

\begin{table*}[t]
\caption{
Comparison with Kling-Foley on the TV2A task across VGGSound-Test, Kling-Audio-Eval, and MovieGen-Audio-Bench.
We report representative metrics including semantic alignment (IB, CLAP), temporal synchronization (Desync),
audio quality (IS), and distribution matching (KL).
CLAP is computed using the MS-CLAP model, IS is based on PANNs, and KL is computed using PaSST features.
The better results between the two methods are highlighted in bold.
}
\centering
\renewcommand{\arraystretch}{1}
\setlength{\tabcolsep}{2.6pt}

\begin{tabular}{p{1.7cm}p{4.1cm}ccccc}
\toprule
Dataset & Method 
& \multicolumn{2}{c}{Semantic alignment} 
& Temporal sync
& Audio quality
& Distribution matching \\

\cmidrule(lr){3-4}
\cmidrule(lr){5-5}
\cmidrule(lr){6-6}
\cmidrule(lr){7-7}

& 
& IB $\uparrow$
& CLAP $\uparrow$
& Desync $\downarrow$
& IS $\uparrow$
& KL $\downarrow$ \\

\midrule

\multirow{2}{*}{\makecell{VGGSound\\Test}}

& Kling-Foley
&  0.30 & 0.31  
&  0.47
&  15.05
&  1.84 \\

& \textbf{ControlFoley (ours)}
& \textbf{0.32} & \textbf{0.36}
& \textbf{0.42}
& \textbf{22.08}
& \textbf{1.43} \\

\midrule

\multirow{2}{*}{\makecell{Kling\\Audio-Eval}}

& Kling-Foley
& 0.22 & 0.37 
& 0.61 
& 6.86 
& \textbf{1.42} \\

& \textbf{ControlFoley (ours)}
& \textbf{0.28} & \textbf{0.38}
& \textbf{0.52}
& \textbf{9.09}
& 2.07 \\

\midrule

\multirow{2}{*}{\makecell{MovieGen\\Audio-Bench}}

& Kling-Foley
& 0.23 & 0.45 
& 0.77 
& 7.64 
& \textbf{2.21} \\

& \textbf{ControlFoley (ours)}
& \textbf{0.27} & \textbf{0.46}
& \textbf{0.69}
& \textbf{8.50}
& 2.38 \\

\bottomrule
\end{tabular}

\label{tab:kling_tv2a}
\end{table*}

\subsection{Comparison with an Industrial Baseline}

To further evaluate the practical competitiveness of our method, 
we compare with a representative industrial system, Kling-Foley ~\cite{wang2025klingfoley}, 
a large-scale multimodal V2A model trained on massive proprietary data. 
Kling-Foley adopts a multimodal diffusion transformer with dedicated modules for 
visual semantic alignment and audio-visual synchronization, 
and serves as a strong industrial baseline. 
We obtain its results via the official API using default settings, 
ensuring all models are evaluated under identical input conditions without manual tuning.

\noindent\textbf{TV2A Results.}
We evaluate on three datasets with different domain characteristics: 
VGGSound-Test (in-domain for our model), 
Kling-Audio-Eval (in-domain for Kling-Foley), 
and MovieGen-Audio-Bench (out-of-domain for both). 

As shown in \cref{tab:kling_tv2a}, our method consistently outperforms Kling-Foley 
in semantic alignment, temporal synchronization, and audio quality across all datasets. 
Notably, this advantage holds not only on our in-domain dataset, but also on Kling-Foley’s own evaluation set, 
as well as the out-of-domain benchmark, demonstrating strong generalization ability.

For example, our method achieves higher CLAP and IB scores while reducing synchronization error, 
and yields significantly better perceptual quality (IS), indicating more accurate and realistic audio generation.

We observe that Kling-Foley performs better on distribution matching (KL), 
which is expected given its large-scale data-driven training paradigm. 
However, our method consistently achieves superior results in alignment, synchronization, 
and perceptual quality, which are more critical for controllable V2A generation. 
These results indicate that explicitly modeling cross-modal control signals 
is important for achieving strong performance in controllable V2A generation, 
complementing large-scale data-driven approaches.

\begin{figure}[h]
\centering
\includegraphics[width=\linewidth]{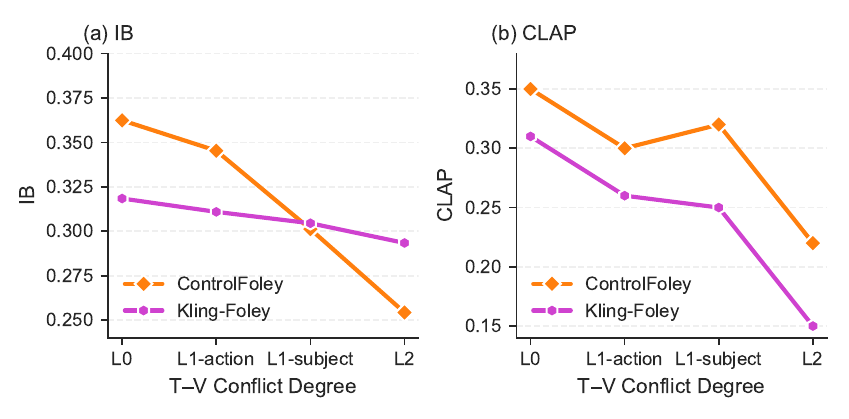}
\caption{
Comparison with Kling-Foley under increasing levels of video-text conflict (TC-V2A).
We report IB (visual consistency) and CLAP (text alignment) across different conflict levels (L0–L2).
Our method exhibits a more flexible trade-off between modalities, with a stronger shift toward textual alignment under conflict.
}
\label{fig:tc_v2a_kling}
\end{figure}

\noindent\textbf{TC-V2A Results.}
We further evaluate controllability under conflicting video-text conditions with increasing levels of T–V conflict.
As shown in \cref{fig:tc_v2a_kling}, both methods exhibit decreasing CLAP scores as the conflict degree increases,
indicating that stronger conflicts make semantic alignment more challenging.

However, a key difference lies in how the two models balance visual and textual modalities.
Our method shows a more pronounced decrease in IB compared to Kling-Foley,
suggesting that it more actively shifts from visual consistency to textual alignment under conflict.
In contrast, Kling-Foley maintains relatively stable IB values,
indicating a stronger bias toward preserving visual consistency even when it conflicts with the text.

As a result, our method consistently achieves higher CLAP scores than Kling-Foley across all conflict levels,
while allowing a larger reduction in IB when necessary.
This behavior demonstrates a more flexible and controllable trade-off between modalities.

The above results suggest that our model better adapts to conflicting conditions by dynamically adjusting modality priorities,
highlighting stronger controllability compared to Kling-Foley.

\noindent\textbf{Summary.}
Overall, the comparison shows that our method consistently outperforms Kling-Foley 
in alignment, synchronization, and perceptual quality, while exhibiting stronger controllability 
under conflicting conditions. 
These results highlight the advantage of explicitly modeling cross-modal control signals, 
which enables more flexible and task-adaptive behavior compared to purely data-driven approaches.

\subsection{User Study}

To complement objective evaluation, we conduct subjective experiments to assess perceptual quality and controllability across three tasks: TV2A, TC-V2A, and AC-V2A.

\begin{figure}[t]
  \centering
  \includegraphics[width=1.0\linewidth]{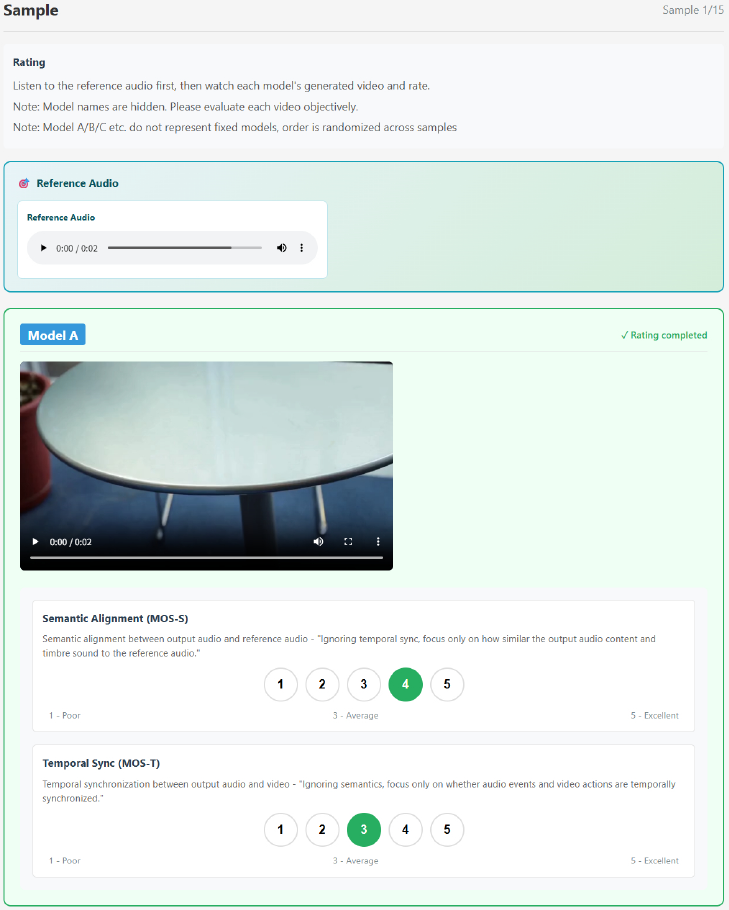}
  \caption{Subjective evaluation page.}
  \label{fig:subjective-evaluation}
\end{figure}

\noindent\textbf{Experimental Setup.} 
We recruit 20 professional participants with relevant audio-visual experience. Before evaluation, detailed instructions are provided for each task and metric. Participants are unaware of model identities, and results for each sample are randomly shuffled to mitigate order bias. The subjective evaluation page is shown in \cref{fig:subjective-evaluation}. 
We note that the user study focuses on comparisons with open-source baselines, while evaluation with closed-source systems (e.g., Kling-Foley) is conducted using objective metrics only due to limited controllability and reproducibility in API-based inference.

\noindent\textbf{Subjective Metrics.}
We adopt Mean Opinion Score (MOS) on a scale from 1 (poor) to 5 (excellent) to evaluate perceptual quality.

For TV2A, we evaluate: 
(i) \textit{MOS-S-AV}, measuring semantic alignment between audio and video; 
(ii) \textit{MOS-T}, assessing temporal synchronization; and 
(iii) \textit{MOS-Q}, reflecting audio quality in terms of clarity, naturalness, and overall listening experience.

For TC-V2A, we use: 
(i) \textit{MOS-S-AT}, evaluating semantic alignment between audio and text; and 
(ii) \textit{MOS-T}.

For AC-V2A, we adopt: 
(i) \textit{MOS-S-AA}, measuring timbre similarity between generated and reference audio; and 
(ii) \textit{MOS-T}.

\noindent\textbf{TV2A Results.}
We randomly select 30 samples from VGGSound-Test, Kling Audio-Eval, and MovieGen Audio-Bench (10 samples each dataset). 
As shown in \cref{tab:user_tv2a}, ControlFoley achieves the best performance across all metrics. It attains the highest MOS-S-AV, indicating superior perceptual alignment. While IB reflects low-level correlation, human perception relies more on semantic consistency. ControlFoley benefits from strong text alignment, which also improves audio-visual alignment in semantically consistent scenarios. Meanwhile, the highest MOS-T aligns with the lowest DeSync values, demonstrating strong temporal consistency. ControlFoley also achieves the highest MOS-Q, consistent with its superior IS, indicating improved audio quality.

\begin{table}[h]
\caption{Subjective results on TV2A.}
\centering
\small
\setlength{\tabcolsep}{5pt}
\begin{tabular}{lccc}
\toprule
Method & MOS-S-AV $\uparrow$ & MOS-T $\uparrow$ & MOS-Q $\uparrow$ \\
\midrule
AudioX & 3.74$\pm$1.02 & 3.34$\pm$1.21 & 3.46$\pm$1.14 \\
Hunyuan & 3.93$\pm$1.17 & 4.10$\pm$0.97 & 3.71$\pm$1.00 \\
MMAudio & 3.89$\pm$1.10 & 3.90$\pm$1.04 & 3.54$\pm$1.00 \\
ThinkSound & 3.45$\pm$1.04 & 3.68$\pm$1.04 & 3.45$\pm$0.89 \\
\textbf{ControlFoley} & \textbf{4.28$\pm$0.87} & \textbf{4.32$\pm$0.82} & \textbf{3.99$\pm$0.83} \\
\bottomrule
\end{tabular}
\label{tab:user_tv2a}
\end{table}

\noindent\textbf{TC-V2A Results.}
We evaluate 30 samples from VGGSound-TVC across different conflict levels (10 samples each for L1-subject, L1-action and L2). 
As shown in \cref{tab:user_tcv2a}, ControlFoley significantly outperforms all baselines in MOS-S-AT, demonstrating strong textual controllability under cross-modal conflict. This is consistent with its superior CLAP performance across increasing conflict levels. In addition, ControlFoley achieves the highest MOS-T, further validating its ability to maintain temporal alignment even under challenging conditions. Interestingly, ThinkSound achieves high MOS-S-AT among baselines. 
This is likely due to its consistently low IB scores, indicating weak reliance on visual information, while maintaining reasonable CLAP performance. 
As a result, its outputs tend to align more closely with text, leading to better perceived text-audio alignment. 
However, this reflects a bias toward text rather than balanced multimodal control, whereas ControlFoley adaptively handles cross-modal conflict.

\begin{table}[h]
\caption{Subjective results on TC-V2A.}
\centering
\small
\setlength{\tabcolsep}{5pt}
\begin{tabular}{lcc}
\toprule
Method & MOS-S-AT $\uparrow$ & MOS-T $\uparrow$ \\
\midrule
AudioX & 2.03$\pm$1.23 & 3.25$\pm$1.18 \\
Hunyuan & 2.19$\pm$1.08 & 3.84$\pm$1.06 \\
MMAudio & 2.00$\pm$1.25 & 3.84$\pm$1.04 \\
ThinkSound & 2.97$\pm$1.36 & 3.86$\pm$1.01 \\
\textbf{ControlFoley} & \textbf{3.85$\pm$1.08} & \textbf{4.02$\pm$1.06} \\
\bottomrule
\end{tabular}
\label{tab:user_tcv2a}
\end{table}

\noindent\textbf{AC-V2A Results.}
We evaluate 30 samples from the Greatest Hits dataset and the AV-VAS dataset (10 samples each dataset). 
As shown in \cref{tab:user_acv2a}, ControlFoley achieves the best MOS-S-AA and MOS-T, consistent with its superior Resemblyzer and DeSync scores. However, MOS-S-AA scores remain relatively low across all methods, indicating that timbre alignment with reference audio is inherently more challenging than semantic alignment with text due to the subtle perceptual characteristics of audio timbre.

\begin{table}[h]
\caption{Subjective results on AC-V2A.}
\centering
\small
\setlength{\tabcolsep}{5pt}
\begin{tabular}{lcc}
\toprule
Method & MOS-S-AA $\uparrow$ & MOS-T $\uparrow$ \\
\midrule
AudioX & 2.49$\pm$1.15 & 3.17$\pm$1.23 \\
CondFoleyGen & 2.86$\pm$1.26 & 3.47$\pm$1.17 \\
\textbf{ControlFoley} & \textbf{2.96$\pm$0.97} & \textbf{3.53$\pm$1.05} \\
\bottomrule
\end{tabular}
\label{tab:user_acv2a}
\end{table}

\noindent\textbf{Discussion.}
Overall, subjective results are highly consistent with objective metrics, validating the effectiveness of our evaluation protocol. ControlFoley consistently achieves strong temporal synchronization across all tasks, highlighting the advantage of the proposed spatio-temporal modeling. The most significant improvement is observed in TC-V2A, demonstrating the effectiveness of our cross-modal conflict handling and modality-aligned training. Meanwhile, results on AC-V2A reveal the intrinsic difficulty of timbre control in V2A task, suggesting an important direction for future research.

\subsection{Ablations}

\begin{table*}[t]
\caption{Ablation study on the REPA loss and its alignment position for the TV2A task on the VGGSound-Test set. 
We compare different configurations by varying whether REPA loss is applied and where it is injected within the DiT blocks (unimodal vs. multimodal, shallow vs. deep layers).  
Best results are highlighted in bold.}
\centering
\renewcommand{\arraystretch}{1.0}
\setlength{\tabcolsep}{2.4pt}

\begin{tabular}{p{5.0cm}ccccc}
\toprule
Settings 
& \multicolumn{2}{c}{Semantic alignment} 
& Temporal sync & Audio quality & Distribution matching\\

\cmidrule(lr){2-3}
\cmidrule(lr){4-4}
\cmidrule(lr){5-5}
\cmidrule(lr){6-6}

& IB $\uparrow$
& CLAP$_{\text{MS}} \uparrow$
& DeSync $\downarrow$
& IS$_{\text{PANNs}} \uparrow$
& KL$_{\text{PaSST}} \downarrow$\\

\midrule

w/o REPA loss
& 0.30 & 0.34
& 0.43
& \textbf{22.47}
& 1.53\\

unimodal, 24th layer
& 0.31 & 0.35
& \textbf{0.42}
& 21.01
& 1.45\\

multimodal, 8th layer
& 0.31 & \textbf{0.36}
& 0.43
& 21.82
& 1.44
\\

\textbf{ours (unimodal, 8th layer)}
& \textbf{0.32} & \textbf{0.36}
& \textbf{0.42}
& 22.08
& \textbf{1.43}\\

\bottomrule
\end{tabular}

\label{tab:ablation_v2a_results}
\end{table*}

To better understand the contribution of each component in ControlFoley, we conduct extensive ablation studies on the key design choices.

\begin{figure}[t]
\centering
\includegraphics[width=\linewidth]{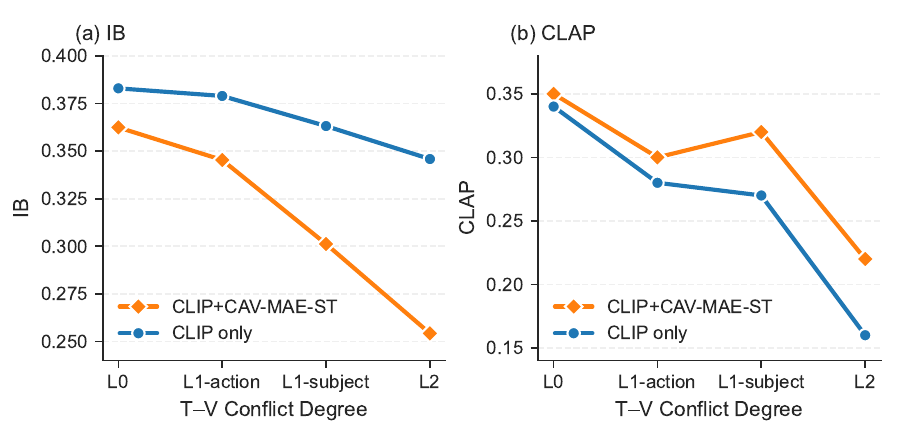}
\caption{Ablation results of joint visual encoding under different T-V conflict degrees. 
We compare the CLIP-only baseline and the proposed joint visual encoding design. 
(a) IB score. (b) CLAP score. 
The meanings of all metrics are consistent with \cref{fig:tc_v2a}.}
\label{fig:ablation_tc_v2a}
\end{figure}

\noindent\textbf{The Effect of Joint Visual Encoding.} 
To investigate the impact of joint visual encoding on textual controllability, we conduct ablation experiments under the TC-V2A setting. 
We compare two configurations: (1) visual features encoded solely by a CLIP encoder, and (2) joint visual encoding using both CLIP and CAV-MAE-ST encoders within our framework.

Experimental results illustrated in \cref{fig:ablation_tc_v2a} show that the joint visual encoding configuration generally outperforms the CLIP-only baseline across different T-V conflict levels. 
The joint encoding design maintains better alignment with textual guidance while more effectively suppressing irrelevant visual information under increasing cross-modal conflict.

We also observe an interesting phenomenon under the no-conflict condition (L0): the CLIP-only configuration achieves slightly higher IB scores than the joint encoding variant. 
This behavior is expected and can be attributed to the shared representation space in the CLIP-only setting. 
Since both visual and textual features are encoded by the same CLIP model, they are naturally projected into a highly aligned semantic space, leading to stronger cross-modal consistency when the modalities are semantically consistent.

However, this tight coupling also introduces limitations under conflicting conditions. 
When the visual and textual inputs become inconsistent, their representations remain strongly entangled, which may intensify modality competition and hinder effective textual control. 
In contrast, the joint encoding design incorporates CAV-MAE-ST to introduce an additional audio-visual representation with different inductive biases, which is less directly aligned with textual semantics. 
This decoupling helps reduce cross-modal interference and improves robustness under increasing T-V conflict.

As shown in \cref{fig:cka}, to verify that joint visual encoding reduces visual modality dominance, we add a CKA (Centered Kernel Alignment) analysis, where higher values mean stronger representation coupling, across VGGSound-TVC conflict levels because L0$\rightarrow$L3 progressively separates text semantics from visual content. High CLIP-Text coupling shows why CLIP visual features can proxy text semantics and make text redundant; low/stable CAV-Text means CAV-MAE-ST does not introduce another text-semantic competitor; low CLIP-CAV indicates complementary visual branches; and Joint-Text below CLIP-Text shows the final joint visual feature is less tightly coupled to text. This supports representation-level mitigation without heuristic arbitration.

\begin{table}[h]
\centering
\setlength{\tabcolsep}{3.0pt}
\caption{Mechanistic CKA across conflict levels.}
\begin{tabular}{lcccc}
\toprule
Level & CLIP-Text & CAV-Text & CLIP-CAV & Joint-Text\\
\midrule
L0 & 0.31 & 0.06 & 0.12 & 0.11\\
L1-sub & 0.24 & 0.06 & 0.13 & 0.10\\
L1-act & 0.30 & 0.06 & 0.14 & 0.11\\
L2 & 0.13 & 0.03 & 0.12 & 0.05\\
L3 & 0.07 & 0.03 & 0.13 & 0.04\\
\bottomrule
\end{tabular}
\label{fig:cka}
\end{table}

Overall, these results suggest a trade-off between alignment strength and robustness to cross-modal conflict. 
By jointly leveraging vision-language and audio-visual representations, our method provides a more balanced and principled solution for controllable text-guided audio generation.

\noindent\textbf{The Effect of Reference Audio Control.} We conduct ablation study on the AC-V2A task to validate our dual-path reference audio conditioning mechanism, evaluating three configurations: (1) removing the semantic conditioning, (2) discarding the timbre conditioning, and (3) retaining both pathways as proposed.

As presented in \cref{tab:ablation_acv2a_results}, our full configuration achieves the best performance in both timbre similarity and temporal synchronization. 
Compared to the setting without semantic conditioning, the full model improves Resemblyzer by 0.25 and reduces DeSync by 0.07, demonstrating that the semantic conditioning branch effectively captures high-level acoustic characteristics while preserving alignment. 

\begin{table}[h]
\caption{Ablation study of the dual-path reference audio conditioning mechanism on the AC-V2A task. 
The two conditioning pathways correspond to global semantic conditioning and global timbre conditioning. 
Best results are highlighted in bold.}
\centering
\renewcommand{\arraystretch}{1.0}
\setlength{\tabcolsep}{3.5pt}

\begin{tabular}{lcc}
\toprule
Settings 
& Timbre sim.
& Temp. sync \\

\cmidrule(lr){2-2}
\cmidrule(lr){3-3}

& Resemblyzer $\uparrow$
& DeSync $\downarrow$ \\

\midrule

w/o semantic conditioning
& 0.56
& 0.92\\

w/o timbre conditioning
& 0.79
& 0.86\\

\textbf{ours (with both conditioning)}
& \textbf{0.81}
& \textbf{0.85}\\

\bottomrule
\end{tabular}

\label{tab:ablation_acv2a_results}
\end{table}

Compared to the setting without timbre conditioning, the full model yields a 0.02 gain in Resemblyzer and a 0.01 reduction in DeSync, indicating that the global timbre pathway further enhances style consistency without disrupting synchronization.  

Overall, these results verify that the combination of global semantic and global timbre conditioning provides complementary information, leading to improved acoustic controllability and temporal alignment for video-to-audio generation.

\noindent\textbf{The Effect of Representation Alignment Training Strategy.}

The final ablation experiment validates the effectiveness of incorporating the REPA loss during training and the optimal choice of its alignment position within the DiT blocks. 
This experiment is conducted on the TV2A task using the VGGSound-Test benchmark. 
We evaluate four configurations: (1) training without REPA loss, (2) applying REPA loss at a deep layer of the unimodal DiT blocks, (3) applying REPA loss within the multimodal DiT blocks, and (4) our proposed design that injects REPA loss at a shallow layer of the unimodal DiT blocks.

As shown in \cref{tab:ablation_v2a_results}, our design achieves the best overall performance. 
Removing the REPA loss leads to degradation in semantic alignment (IB: 0.30 vs. 0.32, CLAP: 0.34 vs. 0.36) and distribution matching (KL: 1.53 vs. 1.43), highlighting its role in preserving semantic consistency between modalities. 
Although the model without REPA achieves the highest audio quality score (IS: 22.47), it suffers from weaker semantic alignment with all modalities.

Applying REPA loss at a deep unimodal layer or within the multimodal transformer results in inferior performance compared to our design, suggesting that early-stage unimodal alignment is more effective for propagating semantic guidance while preserving fine-grained audio details. 
Overall, these results demonstrate that both the presence of REPA loss and its placement are crucial for achieving balanced performance across semantic alignment, temporal synchronization, and audio quality.

\section{Conclusion}

In this paper, we presented ControlFoley, a unified multimodal V2A generation framework that enables precise and robust controllability across text, video, and reference audio. To address the limitations of existing methods, we introduced three key designs: (1) a joint visual encoding paradigm that integrates vision–language and audio–visual representations to enhance textual controllability under cross-modal conflict; (2) a temporal–timbre decoupling strategy that isolates timbre information from reference audio for accurate and interference-free style control; and (3) a modality-robust training scheme with unified multimodal representation alignment (REPA) to ensure stable performance under diverse modality conditions. In addition, we proposed VGGSound-TVC, a new benchmark for systematically evaluating textual controllability under varying degrees of visual-text semantic conflict. Extensive experiments demonstrate that ControlFoley achieves state-of-the-art performance across multiple V2A tasks, while maintaining strong controllability, synchronization, and audio quality.

Despite these promising results, challenges remain for future work. The current training data relies on relatively simple textual annotations, which limits the upper bound of fine-grained semantic control. Notably, the effectiveness of richer or more structured annotations fundamentally depends on the model’s ability to follow textual guidance. Therefore, enhancing textual controllability, as pursued in this work, serves as an important prerequisite for future exploration of more expressive and fine-grained control signals. 

We hope that ControlFoley and the proposed benchmark can serve as a solid foundation for future research on controllable and interactive audio generation.

\bibliographystyle{ACM-Reference-Format}
\bibliography{sample-base}

\clearpage
\appendix
\section*{Appendix}
\addcontentsline{toc}{section}{Appendix}

\section{VGGSound-TVC Construction Prompt}
\label{appendix:propmpt}
\lstset{
    language=Tex,      
    basicstyle=\ttfamily\small, 
    keywordstyle=\color{blue},   
    commentstyle=\color{green!70}, 
    stringstyle=\color{red},  
    numbers=none,           
    numberstyle=\tiny\color{gray}, 
    frame=single,            
    breaklines=true,    
    breakautoindent=false,
    backgroundcolor=\color{gray!5} 
}
\begin{lstlisting}
You are an expert multimodal dataset curator and Foley artist. Your task is to generate rewritten text labels for a video clip in order to construct "Visual-Text Conflic" data for training a flow-matching audio generation model.
The purpose of these labels is to enhance the model's TEXT CONTROL under conditions where visual content and text prompts are partially or strongly conflicting, while preserving realistic audio generation.
CRITICAL RULE:
All rewritten labels MUST preserve the TEMPORAL STRUCTURE of the sound in the video, including rhythm, duration, continuity, and event density. Labels must be derived from the VIDEO CONTENT, not inferred only from text.
--------------------------------------------------
VGGSound label characteristics (STRICTLY FOLLOW):
- Short, lowercase
- No sentences, no punctuation
- One of the following forms:
  1) noun + verb-ing  (e.g., "dog barking", "people whispering")
  2) gerund phrase    (e.g., "chopping wood", "driving buses")
  3) noun phrase     (e.g., "wind noise", "ambulance siren")
--------------------------------------------------
You will be given:
- A video clip (visual input)
- Its raw label: label_L0_raw
  (Note: label_L0_raw may contain multiple comma-separated synonyms)
--------------------------------------------------
Output JSON format (JSON ONLY, no explanation):
{
  "label_L0": "string",
  "label_L1_subject": "string",
  "label_L1_action": "string",
  "label_L2": "string"
}
--------------------------------------------------
Field definitions and requirements:
1. label_L0 (Clean Ground Truth)
- If label_L0_raw contains multiple comma-separated labels,
  SELECT exactly ONE label that best matches the visual content of the video.
- If label_L0_raw contains only one label, copy it directly without modification.
- Do NOT introduce new semantics or change the event type.
- Keep the label concise and in standard VGGSound format
  (e.g., noun + verb-ing, gerund phrase, or noun phrase).
Examples:
- "female speech, woman speaking" -> "woman speaking"
- "engine accelerating, revving, vroom" -> "engine accelerating"
- "wind noise" -> "wind noise"
--------------------------------------------------
2. label_L1_subject (Weak Conflict: Subject Change)
- Change WHO produces the sound.
- Preserve the TEMPORAL STRUCTURE and overall acoustic pattern.
- The new subject must plausibly produce a similar type of sound.
- IMPORTANT:
  When the subject changes, the ACTION WORD must be updated
  to a natural verb for the new subject.
Examples:
- "dog barking" -> "cat meowing"
- "people clapping" -> "audience clapping"
- "car engine idling" -> "truck engine idling"
--------------------------------------------------
3. label_L1_action (Weak Conflict: Action Change)
- Keep the SAME subject.
- Change the action, but STRICTLY preserve the temporal pattern.
Temporal consistency rules:
- Continuous stays continuous
- Rhythmic stays rhythmic
- Transient stays transient
Examples:
- "people clapping" -> "slapping table"
- "dog barking" -> "dog coughing"
- "wind blowing" -> "wind howling"
--------------------------------------------------
4. label_L2 (Moderate Conflict: Semantic Mismatch, Physical Match)
- Change to a DIFFERENT semantic class.
- The sound must remain acoustically and temporally compatible
  with the video's motion and rhythm.
- Visual meaning may conflict, but sound structure must match.
- Think like a Foley artist.
Examples:
- "dog barking" -> "knocking on door"
- "people clapping" -> "hammering nails"
- "horse galloping" -> "knocking on wooden door"
- "paper crumpling" -> "fire crackling"
--------------------------------------------------
Final checks before output:
1. All labels look like valid VGGSound labels.
2. Temporal structure is preserved across all rewritten labels.
3. Output JSON ONLY. No extra text.
\end{lstlisting}

\end{document}